\documentclass{IEEEoj}
\usepackage{cite}
\usepackage{amsmath,amssymb,amsfonts}
\usepackage{algorithmic}
\usepackage{graphicx,color}
\usepackage{textcomp}
\usepackage[T1]{fontenc}
\usepackage{microtype} 
\usepackage{hyperref}
\hypersetup{breaklinks=true}
\usepackage{xurl}
\usepackage{tikz}
\usetikzlibrary{arrows.meta,positioning,shapes.geometric,shapes.arrows}
\def\BibTeX{{\rm B\kern-.05em{\sc i\kern-.025em b}\kern-.08em
    T\kern-.1667em\lower.7ex\hbox{E}\kern-.125emX}}
\AtBeginDocument{\definecolor{ojcolor}{cmyk}{0.93,0.59,0.15,0.02}}
\def\OJlogo{\vspace{-4pt}\hskip-4pt\includegraphics[height=18pt]{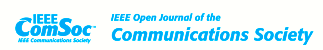}}
\begin{document}
\receiveddate{XX Month, XXXX}
\reviseddate{XX Month, XXXX}
\accepteddate{XX Month, XXXX}
\publisheddate{XX Month, XXXX}
\currentdate{11 January, 2024}
\doiinfo{OJCOMS.2024.011100}

\title{Cybersecurity of High-Altitude Platform Stations: Threat Taxonomy, Attacks and Defenses with Standards Mapping — DDoS Attack Use Case}

\author{Chaouki Hjaiji\IEEEauthorrefmark{1} \IEEEmembership{(Student Member, IEEE)}, Bassem Ouni \IEEEauthorrefmark{2}\IEEEmembership{(Senior Member, IEEE)}, Mohamed-Slim Alouini\IEEEauthorrefmark{3}
\IEEEmembership{(Fellow, IEEE)}}
\affil{Ecole Polytechnique de Tunisie, La Marsa, Tunisia}
\affil{Technology Innovation Institute, Abu Dhabi, United Arab Emirates}
\affil{Computer, Electrical and Mathematical Science and Engineering Division, King Abdullah University of Science and Technology, Thuwal 6900, Saudi Arabia}
\corresp{CORRESPONDING AUTHOR: Chaouki Hjaiji (e-mail: chawki.hjaiji@ept.ucar.tn).}
\markboth{Preparation of Papers for IEEE OPEN JOURNALS}{Author \textit{et al.}}

\begin{abstract}
High-Altitude Platform Stations (HAPS) are emerging stratospheric nodes within non-terrestrial networks. We provide a structured overview of HAPS subsystems and principal communication links, map cybersecurity and privacy exposure across communication, control, and power subsystems, and propose a stratosphere-aware threat taxonomy. We then discuss defenses feasible under HAPS constraints—including encryption and authentication, frequency agility, directional and beam-steered antennas, intrusion detection, secure boot, and software and supply-chain assurance—while highlighting how they align with emerging regulatory and standards guidance. Finally, we report a simulation-based case study using OMNeT++/INET to characterize distributed-denial-of-service (DDoS) impact on service- and control-plane availability, and summarize regulatory and standardization considerations relevant to deployment. We conclude with concrete future research directions. The study is simulation-grounded and intended to inform engineering trade-offs for real-world HAPS deployments rather than serve as an on-air validation.
\end{abstract}


\maketitle

\begin{table*}[!t]
\caption{Summary of main abbreviations}
\label{table:nomenclature}
\setlength{\tabcolsep}{5pt}
\renewcommand{\arraystretch}{1.05}
\centering
\footnotesize
\begin{tabular}{|p{50pt}|p{150pt}|p{50pt}|p{150pt}|}
\hline
    	\textbf{Acronym} & \textbf{Definition} & \textbf{Acronym} & \textbf{Definition} \\
\hline
5G & Fifth Generation & 6G & Sixth Generation \\
ADCS & Attitude Determination and Control System & AI & Artificial Intelligence \\
AODV & Ad hoc On-Demand Distance Vector & APs & Access Points \\
C\&DH & Control and Data Handling & CNN & Convolutional Neural Network \\
COTS & Commercial Off-The-Shelf & CPUs & Central Processing Units \\
c-Si & Crystalline Silicon & CWE & Common Weakness Enumeration \\
DC & Direct Current & DDoS & Distributed Denial-of-Service \\
DL & Deep Learning &  &  \\
DoS & Denial-of-Service &  &  \\
EC & Edge Computing & ECN & Explicit Congestion Notification \\
EO & Earth Observation & FANETs & Flying Ad-hoc Networks \\
FGSM & Fast Gradient Sign Method &  &  \\
FSO & Free-Space Optics & GEO & Geostationary Earth Orbit \\
GRUs & Gated Recurrent Units & GS & Ground Station \\
HAPS & High-Altitude Platform Stations & ICMP & Internet Control Message Protocol \\
IDS & Intrusion Detection System &  &  \\
IoT & Internet of Things & ITU & International Telecommunication Union \\
LEO & Low Earth Orbit & LSTM & Long Short-Term Memory \\
LTE & Long Term Evolution & MIMO & Multiple-Input Multiple-Output \\
ML & Machine Learning & MLP & Multilayer Perceptron \\
NEC & Non-Edge Computing & NOC & Network Operations Center \\
NTN & Non-Terrestrial Networks & OBCS & On-Board Computer System \\
OSS & Open-Source Software & PGD & Projected Gradient Descent \\
RF & Radio Frequency & RFC & Regenerative Fuel Cell(s) \\
RNN & Recurrent Neural Network & SDR & Software Defined Radio \\
SYN & TCP SYN (synchronize) flag &  &  \\
ToS & Type of Service &  &  \\
TT\&C & Telemetry, Tracking, and Control & UAV & Unmanned Aerial Vehicle \\
mmWave & Millimeter Wave \\
TPM & Trusted Platform Module &  &  \\
3GPP & 3rd Generation Partnership Project & IAB & Integrated Access and Backhaul \\
MBS & Macro Base Station & CNS & Communication, Navigation, and Surveillance \\
ICAO & International Civil Aviation Organization & EASA & European Union Aviation Safety Agency \\
FAA & Federal Aviation Administration & ETSI & European Telecommunications Standards Institute \\
ENISA & European Union Agency for Cybersecurity & NIST & National Institute of Standards and Technology \\
ISO/IEC & International Organization for Standardization / International Electrotechnical Commission & SBOM & Software Bill of Materials \\
SSDF & Secure Software Development Framework & RTCA & Radio Technical Commission for Aeronautics \\
EUROCAE & European Organisation for Civil Aviation Equipment & SUPI & Subscription Permanent Identifier \\
NAS & Non-Access Stratum & RRC & Radio Resource Control \\
IMT & International Mobile Telecommunications & QUIC & Quick UDP Internet Connections \\
HTTP/2 & Hypertext Transfer Protocol Version 2 & DSCP & Differentiated Services Code Point \\
QoS & Quality of Service & OMNeT++ & Objective Modular Network Testbed in C++ \\
INET & INET Simulation Framework & HHO & Harris Hawks Optimization \\
DNS & Domain Name System & NTP & Network Time Protocol \\
SSDP & Simple Service Discovery Protocol & TCP & Transmission Control Protocol \\
UDP & User Datagram Protocol &  &  \\
\hline
\end{tabular}
\end{table*}

\section{INTRODUCTION}
\IEEEPARstart{H}{igh}-Altitude Platform Stations (HAPS) operate near 20 km altitude in the stratosphere and can provide wide-area coverage from quasi-stationary positions. Progress in materials, power systems, and autonomous avionics has improved the feasibility for communication and sensing roles. Beyond basic connectivity, envisioned functions include relay and backhaul support, wide-area monitoring, and coordination of aerial systems.

The literature further considers HAPS as compute-enabled platforms and macroscale access points, organized as constellations \cite{vision} to provide high-capacity access, computation offloading, and analytics across diverse environments. HAPS roles in 5G/6G and beyond include:
\begin{enumerate}
    \item Aerial communication hubs, enabling long-distance connectivity without requiring ground-based or offshore relay infrastructure. This capability offers secure, high-speed communication links with satellite networks \cite{using}.
    \item Coordination of Unmanned Aerial Vehicle (UAV) swarms through edge intelligence, offloading computationally intensive tasks while managing extensive sensing and monitoring operations critical for cargo drone applications \cite{vision}.
    \item Extending connectivity and wireless services to urban, suburban, and remote regions, reducing reliance on traditional terrestrial and satellite network infrastructure.
\end{enumerate}
Table \ref{tab:tab1} lists some popular HAPS projects along with their key features.

\begin{table*}
\normalsize
\begin{center}
\caption{Examples of HAPS Projects}
\label{tab:tab1}
\begin{tabular*}{\textwidth}{@{}|p{60pt}|p{80pt}<{\centering}|p{50pt}<{\centering}|p{50pt}<{\centering}|p{202pt}|@{}}
\hline
\textbf{Project} & 
\textbf{Company/ Organization} & 
\textbf{Country} & 
\textbf{Period} & 
\textbf{Description} \\
\hline
ZephyrS \cite{80} & 
Airbus Defense and Space & 
United Kingdom & 
2013-Now & 
It aims to link remote individuals worldwide with characteristics of 100 Mbps of speed, 12 kg of payload, and the ability to fly for 100 days. It is powered by solar energy, as a primary energy source, and batteries in second place, allowing the platform to stay at an altitude of 18 kilometers and above.\\
\hline
Stratobus \cite{82} & 
Thales Alenia Space & 
France & 
2014-Now & 
A designed aircraft with approximately dimensions of 115 m x 34 m and a payload capacity up to 450 kg, able to support 5-year missions. It operates at an altitude of 20 km and with a radius of coverage of 500 km in order to provide 5G communication services. \\
\hline
HAWK30 \cite{83} & 
HAPSMobile & 
Japan & 
2017-Now & 
It has the objective of being the link point for mobile devices, UAVs, and IoT devices worldwide. With characteristics of a 78 m wingspan, 20 km altitude for deployment, and a coverage radius of 100 km, it can operate for multiple months. \\
\hline
PHASA-35 \cite{85} & 
BAE Systems and Prismatic & 
United Kingdom & 
2018-Now & 
It is created for a range of services, such as 5G communications. It can carry up to 15 kg of payload and stay in the air non-stop for a whole year. It can stay at a height of 17-21 km while carrying a payload power of 300-1,000 W, and it has a range of up to 200 km. \\
\hline
Elevate \cite{zero2infinity_elevate} & 
Zero 2 Infinity & 
Spain & 
2009-Now & 
It is a service for transporting payloads to the upper atmosphere to test and validate new HAPS technologies. The STRATOS vehicle can transport up to 100 kg for approximately 24 hours while flying at altitudes ranging from 18 to 22 km. \\
\hline
\end{tabular*}
\end{center}
\end{table*}
HAPS platforms have emerged as critical infrastructure for next-generation communications, operating in the stratosphere to provide wide-area coverage for 5G/6G networks, IoT connectivity, and emergency communications. As these platforms become increasingly autonomous and interconnected, their cybersecurity challenges become paramount. The unique stratospheric environment—characterized by extreme temperature variations, cosmic radiation, and limited physical accessibility—creates attack vectors and operational constraints not present in terrestrial or traditional satellite systems, making HAPS cybersecurity a distinct and urgent research domain.

Recent security incidents have highlighted the vulnerability of aerial systems. The 2022 Viasat KA-SAT cyberattack during the Ukraine conflict demonstrated how satellite communication disruption can have geopolitical implications \cite{viasat2022kasat}. HAPS, operating in the same threat landscape but with different attack surfaces, face similar risks. Unlike satellites, HAPS are accessible for physical maintenance but more vulnerable to ground-based interference due to their lower altitude, creating a unique security profile that requires specialized analysis.

Current HAPS security research focuses primarily on traditional network security approaches without addressing stratosphere-specific challenges. 
This research gap becomes critical as HAPS deployments accelerate for 5G/6G infrastructure and IoT connectivity.

This work presents a comprehensive cybersecurity framework for HAPS platforms, addressing the unique challenges of stratospheric operations. We systematically analyze HAPS architectures, examine cyber threats specific to high-altitude environments, explore regulatory requirements, and propose tailored defense mechanisms. Our approach is validated through experimental DDoS attack simulation, demonstrating both vulnerabilities and the effectiveness of our proposed countermeasures.

\subsection{Research Contributions}
The main contributions of this work are:

\begin{itemize}
    \item \textbf{Comprehensive threat taxonomy:} A systematic classification of cybersecurity threats specific to HAPS platforms, including jamming, replay, system intrusion, data integrity attacks, adversarial attacks on AI systems, DDoS, and supply chain vulnerabilities, organized by affected subsystems (communication payload, TT\&C, control, telemetry, power, and structural components) in Section~\ref{sec:threats}.

    \item \textbf{HAPS-specific vulnerability analysis:} Detailed examination of attack vectors unique to stratospheric environments, including cosmic radiation effects, extreme temperature cycling, and limited maintenance accessibility challenges.

    \item \textbf{Multi-layered defense framework:} Practical security mechanisms tailored for HAPS constraints, including encryption/authentication, frequency hopping with stratospheric implementation considerations, directional antennas, and intrusion detection systems.

    \item \textbf{Experimental validation:} OMNeT++/INET-based DDoS attack simulation demonstrating HAPS vulnerability and validating proposed countermeasures with quantified performance metrics.

    \item \textbf{Regulatory consideration:} Synthesis of cybersecurity requirements from ITU, 3GPP, ICAO, and other relevant standards bodies affecting HAPS deployment and operation.
\end{itemize}

\subsection{Paper Organization}
The remainder of this paper is organized as follows: Section II provides operational overview of HAPS architecture and communication links. Section III reviews related work in HAPS and aerial network security. Section IV presents comprehensive threat modeling and attack analysis specific to HAPS platforms. Section V examines the regulatory and standardization landscape affecting HAPS security. Section VI develops defense mechanisms and countermeasures for HAPS cybersecurity. Section VII validates our approach through experimental case studies using DDoS attack simulation. Section VIII concludes with future research directions.

\section{Operational Overview of HAPS}
        HAPS have emerged as pivotal elements within Non-Terrestrial Networks (NTNs), working alongside satellites and other aerial systems to expand global communication services. These platforms, positioned in the stratosphere, leverage advanced technologies to enhance connectivity across vast geographic regions, and they outperform traditional satellite systems in terms of efficiency and scope. Unlike low-altitude aircraft, these stratospheric platforms, as defined in \cite{5}, are less affected by weather conditions, so they can remain in fixed positions above the Earth for extended periods. Therefore, they can provide similar coverage capacities to Geostationary Earth Orbit (GEO) satellites, with shorter distances. According to the International Telecommunication Union (ITU) \cite{6}, HAPS typically operate at altitudes between 20 and 60 kilometers above the Earth's surface. This position enables low communication delay and minimizes distortion effects from the Earth's atmosphere. Therefore, it allows these platforms to cover large geographic areas with minimal signal degradation, a significant advantage over terrestrial base stations. In reality, HAPS systems can communicate with the ground more efficiently in terms of bandwidth, power, and propagation delay \cite{itu_f1500_2000}. Many commercial projects are looking to use HAPS systems as the primary infrastructure in communication networks due to the intersection of commercial interest and technological advancements in Long Term Evolution (LTE) and 5G networks. As a result, the progress of satellite-connected technology is moving forward with the creation and existence of additional stratospheric nodes \cite{overview}. Mainly, HAPS are not significantly different from Low Earth Orbit (LEO) or GEO satellites in terms of components, but these platforms can maintain continuous operation, adapt to environmental changes, and provide services reliably.
\subsection{HAPS Subsystems} The performance of the HAPS system relies on the collaboration of its core components. Each component not only plays a crucial role on its own but also communicates closely with other parts to ensure smooth and efficient operation. The next paragraph will illustrate the primary components and subsystems of a typical HAPS system. Figure \ref{fig:fig1} depicts the structure of these platforms. We detail each subsystem below.

\begin{enumerate}

    \item \textbf{Flight Control Subsystem}: As noted by Kurt et al. \cite{vision}, the objective of the flight control subsystem is to manage platform stabilization, control its movement, and direct it toward the desired direction. To accomplish its mission, this subsystem relies on sensors to measure altitude and direction, a computing unit to make decisions, and actuators to adjust movement and orientation. HAPS equipped with rotors, powered by the energy subsystem, and guided by the flight control unit, can adapt their path and alignment as needed. Furthermore, the flight control unit handles the connection between the aircraft and the ground station through telemetry, tracking, and command signals. These commands monitor the platform status and facilitate crucial communication between a HAPS and its ground control station \cite{overview}. The system complexity and power requirements depend on the platform mass, operating altitude, and payload needs.

    Quasi-stationary operation is typically specified by station-keeping metrics such as allowable horizontal drift radius (e.g., a few kilometers over 24 hours), root-mean-square (RMS) positional error, and maximum angular rate. Stabilization mechanisms include Global Navigation Satellite System (GNSS)-assisted navigation, inertial sensors, sun/star trackers for attitude reference, and a closed-loop Attitude Determination and Control System (ADCS) used to control spacecraft surfaces, propellers, or vectoring thrusters. For payload pointing, multi-axis gimbals and fine-steering mirrors decouple residual platform motion. Wind shears and stratospheric jets drive slow drift; energy-aware guidance schedules counter-drift maneuvers to maintain beam-spot geometry while minimizing propulsion power.
    Boschetti et al. \cite{overview} found that, for intelligence and Earth Observation (EO) operations, the requirements are much stricter compared to a communications network, which can tolerate minor changes in the position and geometry of the ground radio signal beam spot caused by HAPS movement. In the case of EO operations, specific levels of pointing control accuracy, maneuvering, and stability are required due to the nature of the instrumentation and mission objectives. Furthermore, a HAPS may need to travel to multiple targets across a broad area or track a specific target for multiple days. In these instances, the flight subsystem must be significantly stronger and more intricate.

    \item \textbf{Control and Data Handling (C\&DH) Subsystem}: C\&DH acts as the central control unit of the system that processes the incoming commands from the ground base station to make real-time decisions about navigation and operational adjustments. After receiving control commands from the base station through the TT\&C subsystem, C\&DH routes them to the other HAPS subsystems in order to adapt the platform to the mission requirements. The C\&DH contains components like a flight processor and non-volatile memories needed for software operations, which makes this subsystem a possible target for attacks. The stationary aspect of HAPS systems poses challenges for both the HAPS-related aerial vehicles and ground-based stations in locating their position and determining their path. Therefore, as mentioned in \cite{overview}, a localization and orientation system similar to the one used by geostationary satellites is essential to provide navigation information to the C\&DH subsystem. Sensors like star trackers and sun sensors can be mounted on the HAPS to continuously monitor the position and the altitude of the platform. From another perspective, the integration of machine learning algorithms has been a game-changer, enabling this subsystem to predict and react to operational anomalies or environmental conditions automatically.

    \item \textbf{Power Management Subsystem}: Efficient power management proves critical for continuous HAPS operation, particularly for missions lasting 6-12 months. The power management subsystem coordinates energy distribution across all subsystems while optimizing performance and operational longevity. Contemporary HAPS utilize diverse energy sources, including photovoltaic arrays (e.g., c-Si), hydrogen fuel cells, and advanced battery systems such as lithium-sulfur technologies \cite{vision}. Hydrogen-based regenerative fuel cells (RFC) can store daytime solar energy via electrolysis and reconversion for nocturnal operation, offering attractive specific energy for long-endurance missions.

    The SHARP project \cite{75} demonstrated microwave power beaming using rectenna arrays, achieving radio frequency (RF)-to-direct-current (DC) conversion efficiency, though safety concerns limit practical deployment due to radiation exposure risks. Current commercial HAPS designs primarily integrate hybrid solar-battery systems, where photovoltaic arrays generate power during daylight while charging battery banks for nocturnal operations.

    Stratospheric power systems face unique challenges: solar irradiance varies with altitude, while extreme temperatures affect battery performance. Maximum Power Point Tracking (MPPT) efficiency degrades under rapid temperature cycling. Advanced thermal management using phase-change materials helps maintain battery temperatures within optimal operating ranges, extending cycle life significantly.

    Energy budgets for typical communication HAPS must balance payload operations, propulsion/station-keeping, flight control and navigation, and housekeeping functions. Mission planning algorithms optimize compute/communication duty cycles with solar availability, implementing load shedding protocols during energy deficit conditions to ensure extended autonomous operation.
    \item \textbf{Communication Payload Subsystem}: As noted by Boschetti et al. \cite{overview}, the communication payload subsystem is used to establish connections between the HAPS platform and other entities like ground base stations, satellites, and aerial vehicles. It manages both uplink and downlink communications, ensuring data flow to and from the ground stations and end-users. This subsystem consists of antennas for communication links and a variety of processors that handle reception and transmission tasks.

        Recent advancements in adaptive antenna technology, such as beamforming and Multiple-Input Multiple-Output (MIMO), enable HAPS platforms to dynamically adjust their radiation patterns. This optimization improves signal quality, mitigates interference, and adapts to varying network conditions, user mobility, and environmental factors. Adaptive antennas play a crucial role in maintaining robust communication links and maximizing spectral efficiency for HAPS-based networks.

        The deployment of HAPS impacts both FSO and RF communication systems. Variations in HAPS altitude directly affect RF communication by altering the link budget, path loss, and coverage area. Higher altitudes can extend coverage but may increase propagation losses. For FSO systems, the hovering and movement of HAPS platforms introduce pointing errors and angle-of-arrival (AOA) fluctuations, which can degrade alignment and signal reliability. These factors must be considered in the design and operation of HAPS communication payloads to ensure consistent performance under diverse atmospheric and operational conditions.

    \item \textbf{Telemetry, Tracking, and Control (TT\&C) Subsystem}: Due to the specific operational environment of HAPS platforms, maintaining a continuous connection to the ground control station is challenging. Therefore, the TT\&C subsystem consists of multiple antennas and processors that ensure connectivity between HAPS and the control ground station and ensure the transmission of critical information and control commands from the ground to HAPS, especially to the C\&DH subsystem. Many requirements must be considered during TT\&C subsystem design for efficient data transmission, such as data rate, distance, and operational frequency. Meanwhile, the RF power and antenna design of this subsystem are less complicated and constrained compared to satellites.

        The TT\&C subsystem must be designed to meet specific data rate requirements for reliable command and telemetry exchange, with coverage range tailored to the operational altitude and mission profile of HAPS. Operational frequency selection is critical: TT\&C links often operate in dedicated bands (e.g., S-band, X-band) separate from payload data transmission frequencies to avoid interference and ensure robust control, though in some cases co-frequency operation may be considered for simplicity. The choice of frequency, data rate, and coverage range directly impacts the reliability and cybersecurity of HAPS operations, making careful subsystem design essential.

    \item \textbf{Structural Components}: The integrity of the physical structure of HAPS is essential for enduring the stratospheric conditions. So, the structure components of HAPS must be lightweight and robust enough to support all internal components while resisting high-altitude conditions, extreme temperatures, and radiation levels. Boschetti et al.\cite{overview} identified the three main structural components common to aircraft: envelope, parachute, and gondola.\\
    \textbf{Envelope}: The envelope is a thin plastic membrane, such as polyethylene, able to encompass a gas lighter than air. This envelope must endure the harsh conditions of the stratosphere like temperature fluctuations and ultraviolet radiation. The HAPS requires a gas tank filled with a gas capable of making the platform reach its operational altitude. Therefore, a gas such as helium can cause the envelope to expand, allowing the system to reach extremely high altitudes.\\
    \textbf{Parachute}: Parachutes serve as a protective measure for a safe off-landing of the HAPS platform at the end of the mission or in case of emergency. It protects the platform structure and payload during the ballooning phase.\\
    \textbf{Gondola}: The gondola is a metallic structure attached to the envelope and encompasses the different components of the HAPS platform. Similar to a satellite bus, it provides protection and houses all the HAPS subsystems, antennas, and weights. The design of the gondola takes into consideration environmental conditions, anticipated parachute shock, and the impact on the ground when landing. Additionally, the design ensures the simple recovery of instruments and systems with limited tools in distant locations.
    
    \begin{figure}[ht!] 
        \centering
        \includegraphics[width=.5\textwidth]{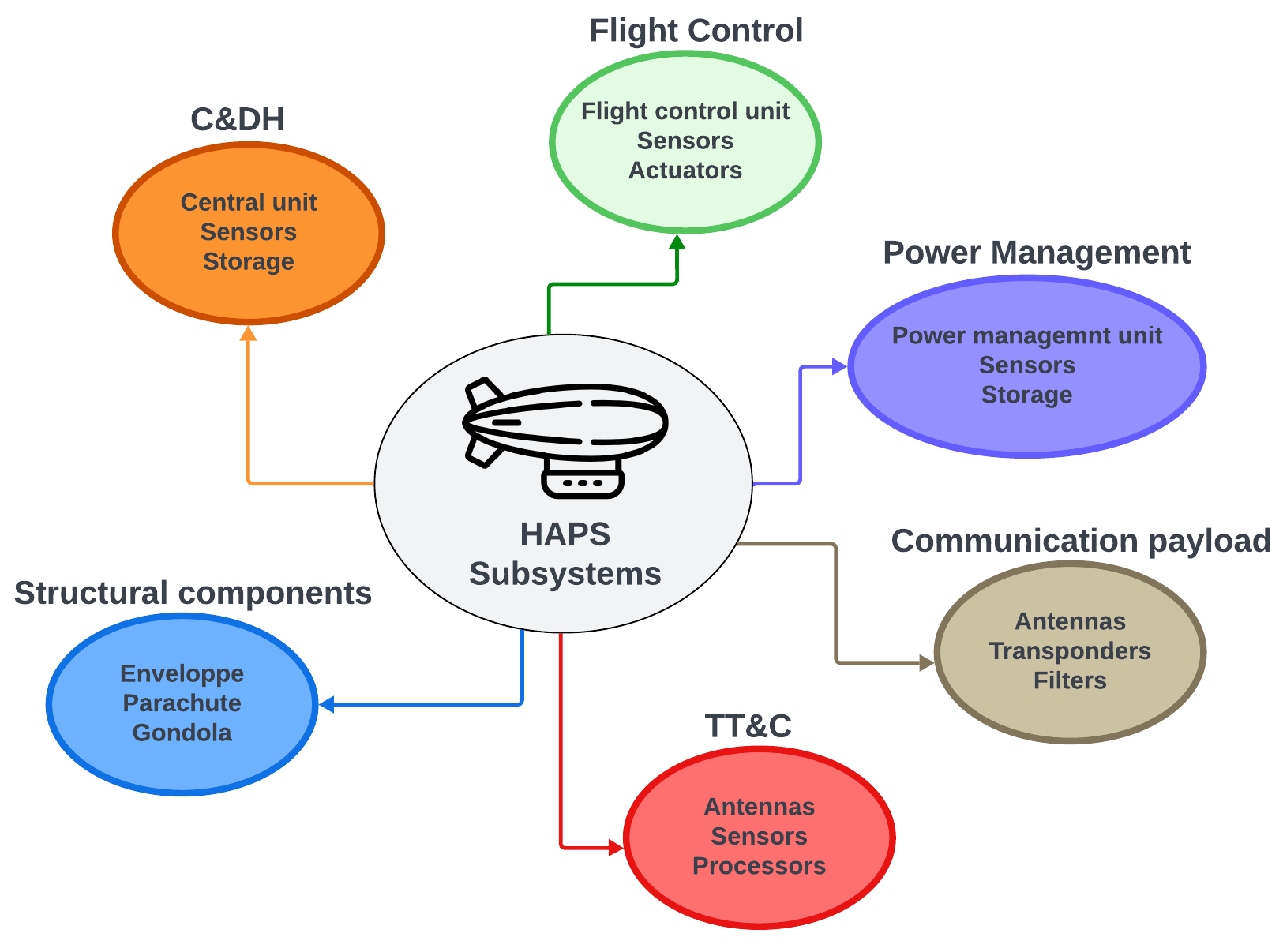} 
        \caption{HAPS Subsystems}
        \label{fig:fig1}
    \end{figure}
\end{enumerate}

\subsection{HAPS Communication Links and Network Architectures}
HAPS are considered NTN platforms, according to the International Telecommunications Union \cite{itu_f1500_2000}, and are able to provide widespread wireless connectivity in areas where it is too expensive or difficult to install fiber optic cables, like remote rural locations, oceans, and mountainous regions. Moreover, NTNs can expand current ground networks, increasing their potential and decreasing communication delays. In this section, we outline the main communication links between HAPS and other NTN platforms like satellites and UAVs. Following this, we discuss the potential network structures in which HAPS play vital roles in upcoming NTNs, such as ad-hoc, cell-free, and integrated access and backhaul configurations \cite{haps}. To highlight the importance of HAPS in NTNs, we offer in-depth insights on using HAPS in these proposed designs with the most suitable communication links, showcasing HAPS' ability to connect NTNs and their adaptability through various roles, such as serving as UAV network centers, satellite relays, and ground network extensions, depending on the specific architecture.\\
HAPS interconnects the NTNs nexus by establishing different communication links between these platforms and other NTN platforms such:\vspace{.2cm}\\
\begin{sloppypar}
                	\textbf{FSO vs. RF Communication: Key Trade-offs}: While this work assumes familiarity with FSO and RF technologies, a brief summary clarifies architectural choices. FSO links offer very high-throughput, license-free operation with narrow beams yielding inherent spatial security and low interference; however, they are highly sensitive to weather (fog, haze, precipitation), scintillation, and require sub-milliradian pointing stability. RF links (microwave/\allowbreak mmWave) are more weather-resilient and tolerant to pointing error, support non-line-of-sight in some bands, and simplify network acquisition and handover, but face spectrum licensing constraints, wider beams (more interference), and typically lower peak spectral efficiency than FSO. Energy-wise, FSO transmitters can be more power-efficient at very high data rates due to tight optical beams, whereas RF systems may require higher energy to close long links—especially at mmWave.
\end{sloppypar}

In practice, HAPS networks adopt hybrid FSO/\allowbreak RF designs: FSO for backhaul under favorable atmospheric conditions and adaptive fallbacks to RF during impairments. Cross-layer controllers can monitor link quality (e.g., scintillation index, visibility) and switch modulation/coding or route traffic over RF failover paths. Mechanical or electro-optical beam stabilization (fine-steering mirrors, gimbals) and adaptive optics mitigate FSO pointing error, while phased-array beamforming improves RF robustness and energy efficiency.

\begin{figure}[ht!] 
        \centering
        \includegraphics[width=0.5\textwidth]{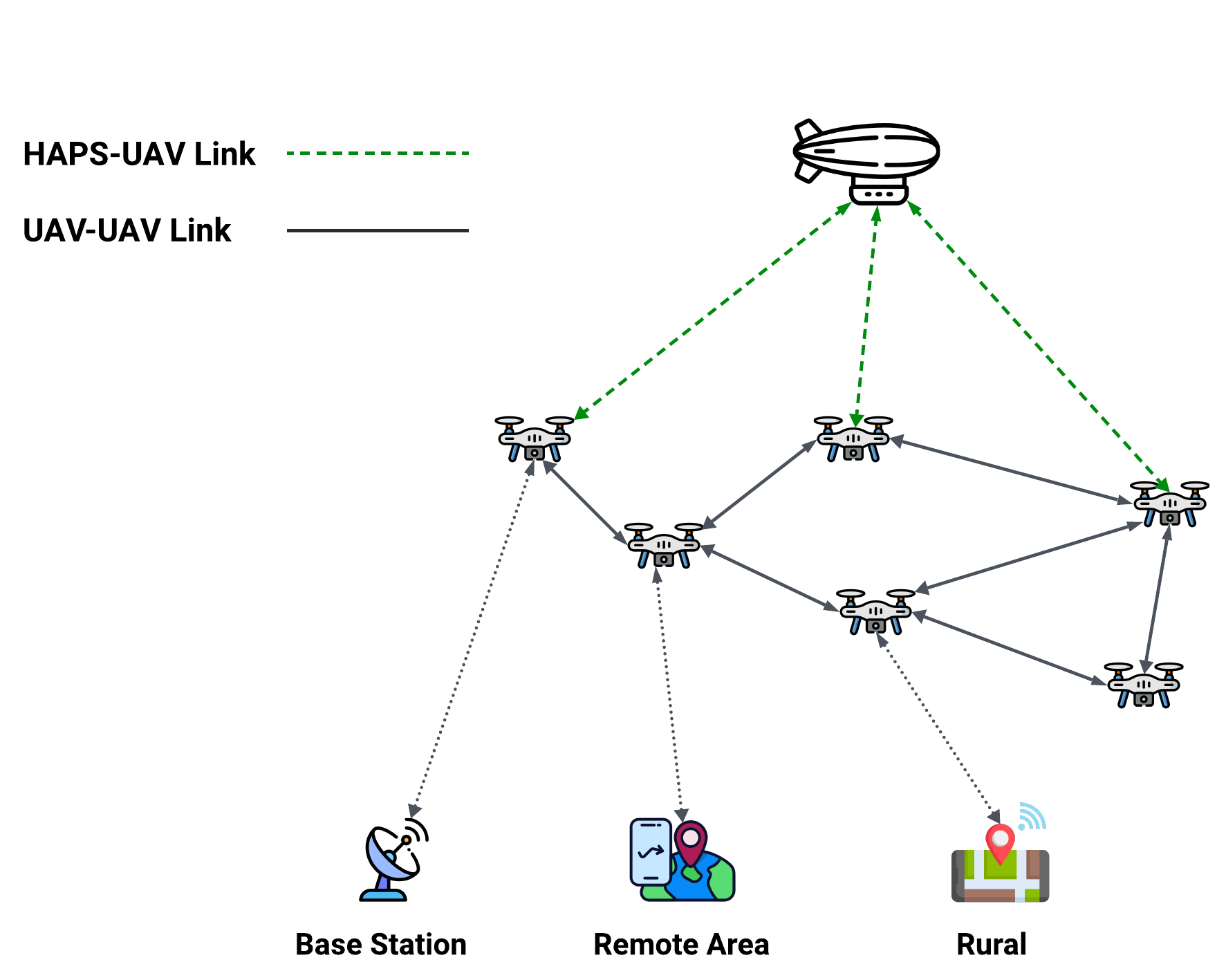}
        \caption{HAPS-Centered Ad-hoc Network Configuration}
        \label{Adhoc}
        
\end{figure}

\begin{figure}[ht!] 
    \centering
    \includegraphics[width=0.4\textwidth]{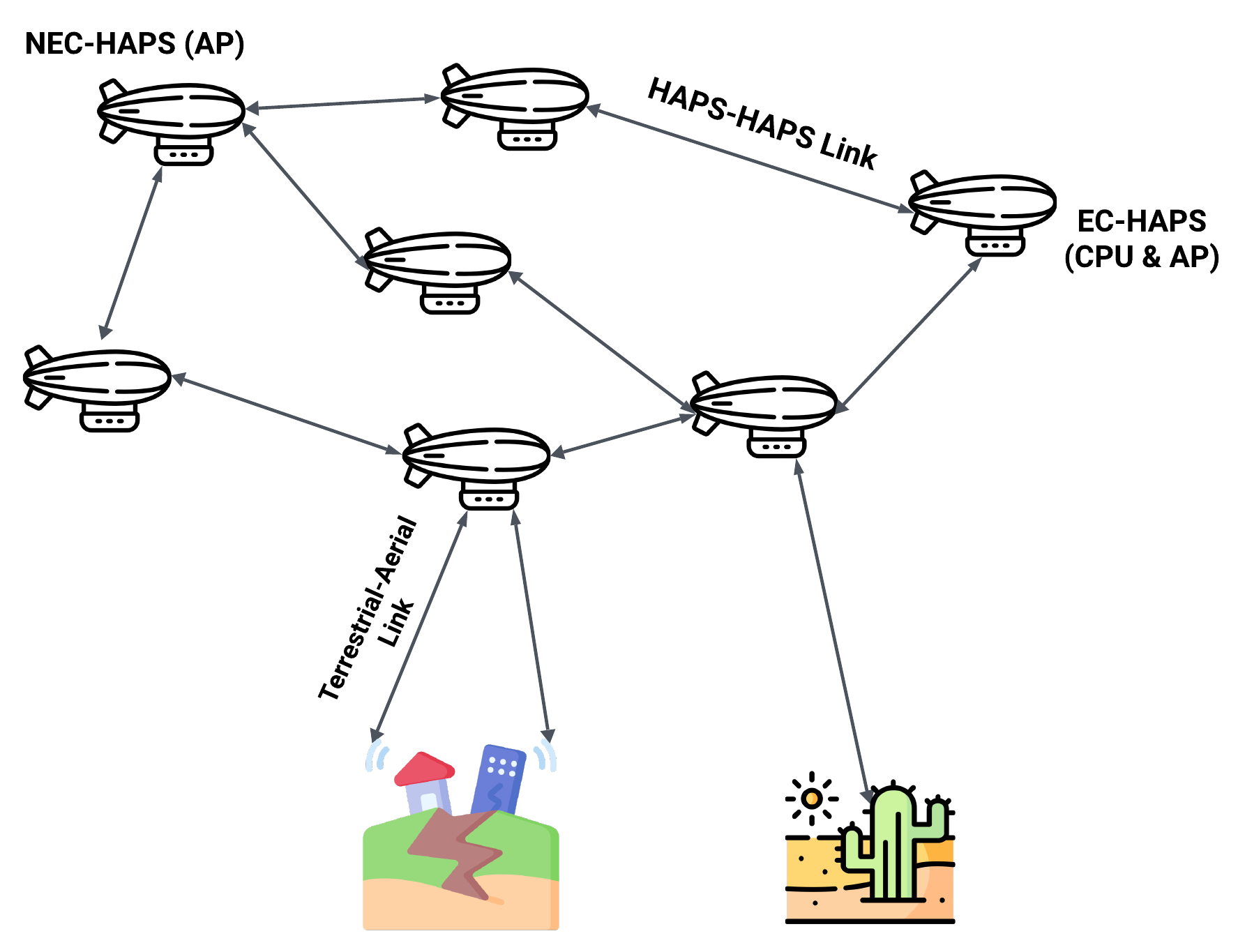}
    \caption{HAPS-Based Cell-Free Network Configuration}
    \label{cell-free}
    
\end{figure}

\begin{figure}[ht!] 
    \centering
    \includegraphics[width=0.5\textwidth]{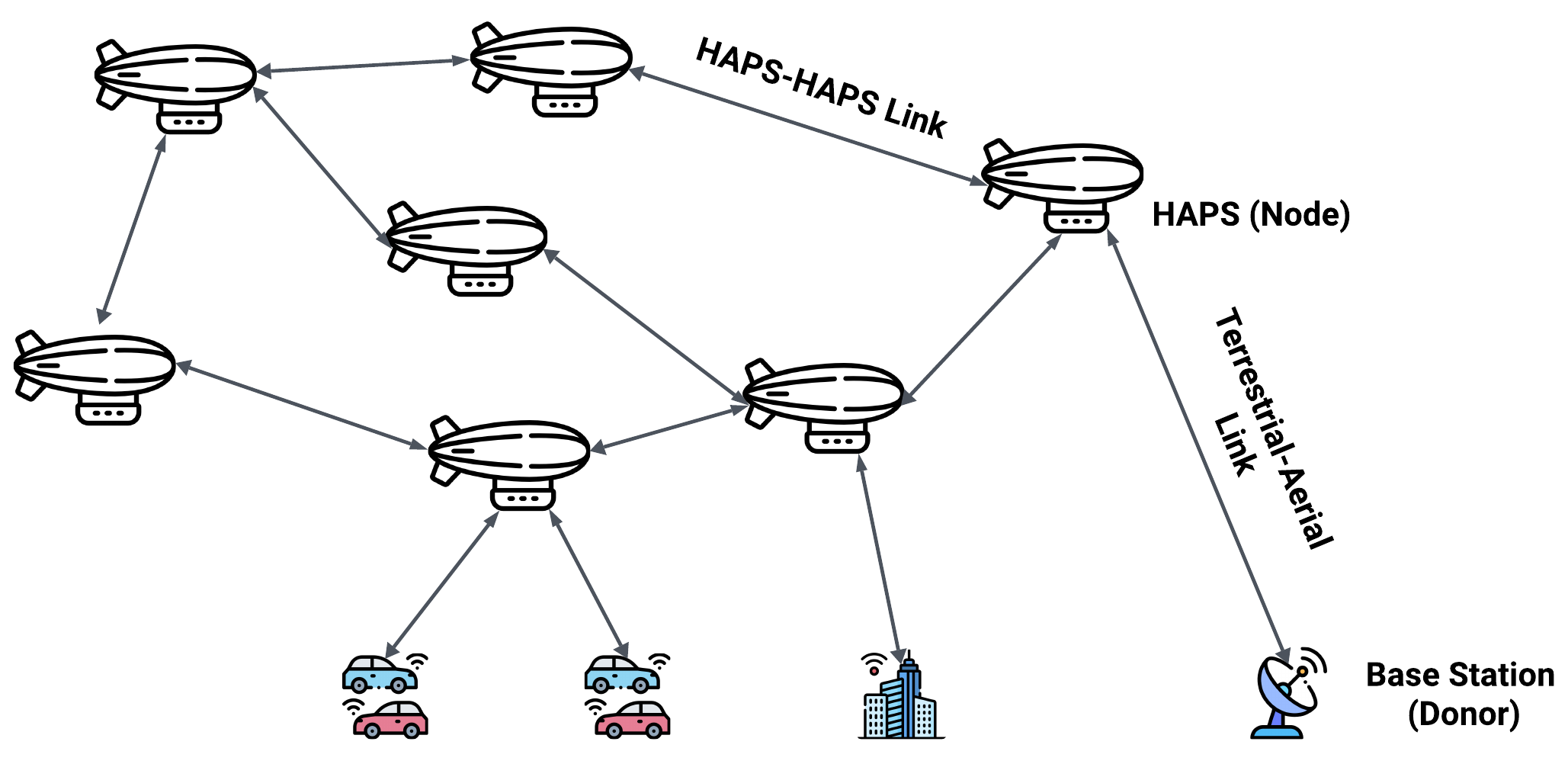}
    \caption{HAPS-Aided Integrated Access and Backhaul Configuration}
    \label{IAB}
    
\end{figure}

\textbf{HAPS-HAPS Link}: These kinds of links are used for long-distance backhaul purposes, such as in emergencies and remote monitoring, where they need a high-speed transmission rate. The popular used frequency band in HAPS to HAPS links is FSO regarding its high data rates and minimal latency. However, this frequency band requires an accurate alignment between transmitter and receiver over long distances, therefore advanced beam steering and tracking methods can be applied or a combination of FSO signals with other technologies such as millimeter wave (mmWave) can serve as a backup communication link \cite{haps}.\\
\textbf{HAPS-Satellite Link}: HAPS can connect with LEO satellites to extend internet coverage to distant areas or serve as a relay hub between satellites and NTNs platforms or end users to improve communication performance. Similar to the HAPS-HAPS links, HAPS to satellite links use high-frequency radio communications and FSO signals as communication methods for establishing connectivity between NTNs and satellites. Maintaining continuous connections between moving satellites and HAPS platforms is challenging, especially with the increasing distance, which results in a significant frequency shift and signal attenuation due to the Doppler effect. To tackle this issue, the use of adaptive optics for FSO signal adjustment could make the laser beams more pointed. As mentioned in \cite{haps}, in order to improve high-frequency radio communication, strategies like space-time coding and beamforming can help reduce signal interference caused by satellite movement.\\
\textbf{HAPS-UAV Link}: HAPS to UAVs connections are generally used in tasks like video surveillance in the military sector or in controlling crops in agriculture \cite{haps}. The established links between HAPS and UAVs via FSO signals could be affected by atmosphere conditions, especially while crossing both the troposphere and stratosphere, which results in energy loss for high-frequency radio signals, affecting by that communication quality. So, using a combined communication systems with integration of FSO and RF signals can offer more reliable and high-speed communication solution.\\
In what follows, we examine the possible network configurations where HAPS is essential for future NTNs, such as ad-hoc, cell-free, and integrated access and backhaul configurations:\vspace{.2cm}\\
    \textbf{HAPS-Centered Ad-hoc Network}: In this network configuration, HAPS serves as the central processing unit of a specific UAV network called Flying Ad-hoc Networks (FANETs). In fact, FANETs made up of multiple UAVs linked together can operate in remote areas and disaster zones. UAVs often need to communicate with nearby UAVs in order to plan the anticipated flight paths and prevent potential crashes. Since the coordination between all UAVs is challenging within FANETs—especially with equal treatment of nodes without distinctions and the fact that any UAV position could impact the entire network—HAPS have been proposed as a solution, with computational abilities to handle this task. In a network configuration centered on HAPS, HAPS must create links with all UAVs, and through these links, UAVs can exchange data within HAPS-focused ad-hoc networks from remote and rural zones to the base station, as shown in Figure \ref{Adhoc}. Therefore, in a HAPS-centered network, UAVs are efficiently managed and routed with generally close connections, resulting in minimal communication lag and low energy usage. The communication link between HAPS and UAVs faces challenges related to reliability and security. To address these challenges, enhancing communication reliability can be achieved through a multi-path transmission system, while implementing encryption can enhance communication safety.
\\
    \textbf{HAPS-Based Cell-Free Network}: This configuration is characterized by HAPS platforms denoted as non-edge computing (NEC)-HAPS and other HAPS platforms that serve as edge computing (EC)-HAPS, similar to a cell-free network composed of central processing units (CPUs) and access points (APs), as illustrated by Figure \ref{cell-free}. Typically, NEC-HAPS are aerial APs created for signal reception and transmission, while EC-HAPS act like edge servers that enable the execution of various computing tasks like extensive data processing, signal detection, and information integration. Furthermore, efficient collaboration among HAPS could reduce the cost of covering a specific area. Within this system, security concerns may arise from edge devices like cyber attacks and hacker breaches. Hence, it is essential to introduce security measures like encrypted communication and identity authentication to guarantee network security.
\\
	\textbf{HAPS-Aided Integrated Access and Backhaul (IAB) Architecture}: Because spectrum resources are limited, 5G introduced IAB configuration, which partitions spectrum between access and backhaul links and enables hierarchical topologies \cite{13}. In the HAPS-assisted IAB architecture of Figure \ref{IAB}, devices operate either as IAB donors or IAB nodes. IAB donors are typically terrestrial equipment such as fiber-connected macro base stations (MBSs) that interface the radio network with the core network. IAB nodes are NTN elements that extend terrestrial coverage via wireless backhaul. A common deployment uses a lower layer of IAB donors and one or more upper layers of IAB nodes, which may include HAPS platforms.

\section{Related Work}
HAPS cybersecurity research intersects multiple domains, including aerial vehicle security, satellite communications, and stratospheric platform engineering. This section positions our work within existing literature while identifying key research gaps that our proposed framework addresses.

\subsection{HAPS Communication Security Challenges}
Recent research has focused on securing high-altitude platform communications through advanced optical and RF systems. Phuchortham and Sabit \cite{phuchortham2025survey} provided a comprehensive survey on FSO communication with RF backup for aerial platforms, examining both theoretical models and real-world applications. Their work addresses the growing demands on network capacity from sensor devices and the limitations of RF technology in densely populated areas, proposing hybrid FSO/RF systems as a solution.

The study categorizes hybrid systems into three mechanisms: hard switching, soft switching, and relay-based approaches, with particular emphasis on multi-platform applications including UAVs, HAPS, and satellites. Their analysis highlights that while FSO offers high data rates and secure communication similar to fiber optics, it remains highly susceptible to atmospheric turbulence and conditions such as fog and smoke. The combination of RF and FSO strengths enables enhanced network reliability for seamless communication in dynamic environments, addressing key challenges in stratospheric platform operations.

However, their work focuses primarily on communication reliability and performance optimization rather than cybersecurity threats specific to stratospheric environments, highlighting a significant gap in comprehensive security frameworks for HAPS operations.

Complementing this communication-focused research, Liu and Wang \cite{liu2025security} analyzed security vulnerabilities in two-factor privacy-preserving protocols for efficient authentication in Internet of Vehicles networks, establishing important security considerations for vehicular network operations. Their work identified authentication weaknesses and proposed enhanced security mechanisms using elliptic curve cryptography. Building on this foundation, Karapantazis and Pavlidou \cite{karapantazis2005broadband} provided a comprehensive survey of broadband communications via high-altitude platforms, examining security challenges across multiple communication scenarios, including cellular coverage, broadband access, and disaster recovery communications.

\subsection{Aerial Network Security Frameworks}
The security of aerial communication networks has received increasing attention as these platforms become critical infrastructure components. Shafique et al. \cite{shafique2021security} provided a comprehensive security analysis of UAV communication networks, identifying key vulnerabilities in aerial-to-ground and aerial-to-aerial communications that are directly applicable to HAPS security considerations. Their framework addresses authentication challenges, secure routing protocols, and intrusion detection mechanisms specifically designed for mobile aerial platforms.

However, their work primarily focuses on low-altitude UAV networks and does not address the unique stratospheric environment challenges faced by HAPS platforms, including extended operational periods, extreme environmental conditions, and limited maintenance accessibility. This limitation highlights the need for stratosphere-specific security frameworks.

Building on aerial network security foundations, Wibawa \cite{wibawa2021security} conducted an extensive review of security architectures for wireless network traffic handoff systems, examining authentication protocols and attack vectors that are relevant to aerial platforms. His analysis covers security threats in wireless handoff processes, including man-in-the-middle attacks and denial-of-service (DoS) attacks that are particularly applicable to HAPS networks operating in the stratosphere.

The evolution of aerial network security has been further advanced by Xie \cite{xie2024security}, who analyzed security vulnerabilities in three-factor authentication schemes for 5G wireless sensor networks in IoT systems. Their work addresses authentication protocol weaknesses and security considerations in next-generation wireless networks, examining vulnerabilities such as user impersonation attacks and sensor node capture attacks. While focused on 5G wireless sensor network (WSN) security analysis, their findings provide valuable insights for HAPS platforms that must implement robust authentication mechanisms while maintaining security in dynamic aerial environments.

\textbf{Stratospheric Environment Specificity}: Existing aerial vehicle security research primarily focuses on UAVs operating below 10 km altitude, failing to address stratospheric-specific challenges, including cosmic radiation effects on electronics, extreme temperature cycling, and limited physical accessibility for maintenance and updates. The unique operational environment of HAPS platforms introduces vulnerabilities not addressed by traditional terrestrial or low-altitude aerial security frameworks.

\textbf{Multi-Domain Attack Scenarios}: Current literature lacks a comprehensive analysis of multi-altitude attack scenarios where adversaries coordinate across terrestrial, aerial, and space domains. The strategic positioning of HAPS platforms creates unique attack vectors that span multiple operational domains, requiring coordinated defense mechanisms not addressed by existing single-domain security solutions.

\textbf{Regulatory Integration Challenges}: Current security frameworks often overlook the complex regulatory environment governing stratospheric operations, including ITU spectrum coordination, aviation safety requirements, and international airspace regulations. This regulatory complexity creates security implementation challenges that are not adequately addressed in existing research.

\textbf{Real-World Deployment Validation}: Several works relies on simulation or controlled environments, with limited validation under actual atmospheric conditions and operational constraints specific to stratospheric platforms. The harsh stratospheric environment presents unique challenges for security system deployment and operation that require specialized consideration.

Despite growing interest in aerial communication systems, several critical security gaps persist in the stratospheric domain.

\subsection{DoS and DDoS Attacks on Aerial Platforms}
DoS attacks represent a significant threat vector for aerial communication platforms due to their critical role in communication infrastructure. Alashhab et al. \cite{alashhab2022ddos} conducted a comprehensive survey of DDoS attacks against cloud computing environments, providing foundational techniques applicable to HAPS network security. Their analysis covers various DDoS attack patterns and mitigation strategies, though adaptation to stratospheric environments requires consideration of unique propagation characteristics and resource constraints.

The challenge of securing aerial platforms against distributed attacks has been addressed by Mo et al. \cite{xu2025practical}, who proposed a practical two-factor authentication protocol for real-time data access in wireless sensor networks. Their approach utilizes robust cryptographic techniques to ensure secure authentication and key agreement, providing foundational security methodologies applicable to multi-HAPS constellation security. Their work demonstrates significant improvements in authentication efficiency and security resilience, offering valuable insights for the unique characteristics of stratospheric HAPS deployment scenarios.

\subsection{Authentication and Key Management in Aerial Networks}
Secure authentication and key management present unique challenges in aerial networks due to mobility, intermittent connectivity, and resource constraints. Wazid et al. \cite{wazid2020authenticated} design a lightweight mutual-authentication and session key agreement protocol for UAV-enabled wireless sensor networks, with formal security analysis (e.g., AVISPA) and performance evaluation via network simulation. The approach is relevant to HAPS when a platform mediates access for ground sensors or UAV clusters; practical adoption would need adjustments for multi-month missions, intermittent backhaul, and constrained on-board compute/storage.

Physical-layer security (PLS) complements upper-layer authentication by improving resilience to eavesdropping at the waveform/link level. Lei et al. \cite{lei2018secrecy} derive closed-form expressions for average secrecy capacity under Wyner’s wiretap model over $\alpha$–$\mu$ fading and analyze high-SNR scaling laws. Although not UAV/HAPS-specific, these results apply to non-terrestrial and aerial links modeled by $\alpha$–$\mu$ fading and can inform HAPS link budgets, antenna patterns, and power allocation in contested environments.



\subsection{Machine Learning for Aerial Network Security}
ML offers adaptive detection and response capabilities for aerial networks, but HAPS deployments face unique constraints: tight power budgets, radiation-induced soft errors, heterogeneous links, and intermittent backhaul that favor lightweight on-board inference with off-platform (ground/cloud) training and model management \cite{vision}. At the same time, ML also enables more capable attacks. Shi et al. \cite{shi2021generative} demonstrate a physical-layer spoofing attack using a generative adversarial network (GAN) “over the air” that crafts synthetic RF signals which mimic waveform, channel, and hardware effects to evade signal authentication, underscoring the need for robustness beyond statistical fingerprints.

Given the absence of public HAPS-specific attack corpora, practical pipelines should disclose dataset provenance and class imbalance and prioritize: (i) low-footprint, explainable anomaly scoring on-board; (ii) secure, signed model updates and periodic integrity checks; and (iii) evaluation that includes operational metrics (energy per inference, latency) alongside detection quality. In short, ML should augment—rather than replace—deterministic safeguards (authentication, rate limiting, secure boot), with gradual adoption aligned to HAPS resource envelopes \cite{vision}.

\subsection{Threat Landscape Evolution}

The evolving threat landscape for high-altitude platforms presents several emerging challenges:

\textbf{Advanced Persistent Threats}: Threat actors increasingly target critical infrastructure, including stratospheric communication platforms that serve as key nodes in global communication networks. The strategic value of HAPS platforms makes them attractive targets for sophisticated adversaries \cite{enisa_supply_chain_2021,nistir_8270_2020}.

\textbf{Supply Chain Vulnerabilities}: The complexity of the development and deployment of the HAPS platform creates numerous opportunities for supply chain attacks, from hardware tampering to software backdoors introduced during the manufacturing and integration process \cite{nist_sp_800_161r1_2022,enisa_supply_chain_2021}.

\textbf{Insider Threats}: The limited operational personnel and extended deployment cycles of HAPS platforms create unique insider threat scenarios that require specialized detection and mitigation strategies \cite{cert_insider_guide_2016,cisa_insider_mitigation_2022}.

\textbf{Environmental Attack Vectors}: The stratospheric environment itself can be weaponized through targeted atmospheric interference, electromagnetic attacks, or exploitation of cosmic radiation effects on platform electronics \cite{rtca_do_160g_2010,iec_62396_2017}.

\section{Potential Cybersecurity Threats}
\label{sec:threats}
\begin{figure}[htbp]
    \centering
    \includegraphics[width=.5\textwidth]{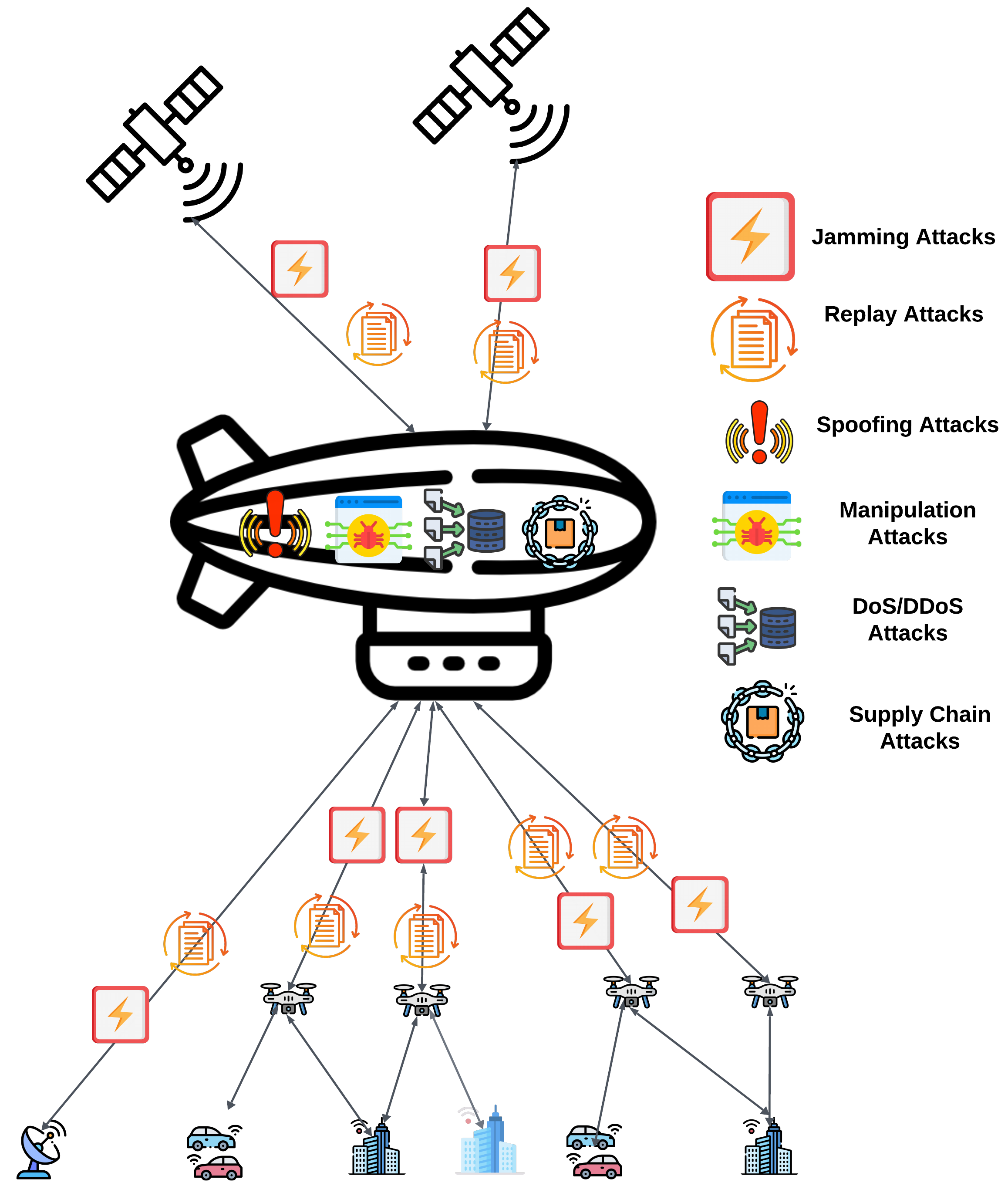}
    \caption{HAPS common cyber threats}
    \label{fig:threats}
\end{figure}

\begin{sloppypar}
HAPS systems, which play crucial roles in communication networks, are vulnerable to a range of cybersecurity threats that could compromise their operations and the integrity of the data they process.
The primary cybersecurity threats facing HAPS, as shown in Figure \ref{fig:threats}, include \textit{jamming and replay attacks} that target the \textit{HAPS communication links}, aiming to disrupt the platform connections within the network. Also, \textit{spoofing attacks}, leading to unauthorized access, affect the command part of the HAPS, mainly the \textit{control subsystem}. Adding to that, the embedded sensors within HAPS platform in \textit{telemetry subsystem} are susceptible to a type of attack, \textit{manipulation attacks}, that would alter and falsify the output data of these devices. \textit{DDoS attacks} would overwhelm the \textit{communication payload subsystem} with floods of requests consuming the HAPS resources. On the other hand, \textit{supply chain attacks} can damage these aircraft by the intrusion of malicious software and hardware components in the HAPS system during its design process.
\end{sloppypar}
As part of our contribution, Table~\ref{tab:threat_taxonomy} formalizes our HAPS threat taxonomy by mapping major threat classes to affected subsystems and primary impacts.

\begin{table}[!t]
\caption{HAPS Threat Taxonomy Mapped to Affected Subsystems}
\label{tab:threat_taxonomy}
\centering
\footnotesize
\setlength{\tabcolsep}{3.5pt}
\renewcommand{\arraystretch}{1.1}
\resizebox{\columnwidth}{!}{%
\begin{tabular}{|p{80pt}|p{95pt}|p{70pt}|}
\hline
    \textbf{Threat Class} & \textbf{Primary Target Subsystems} & \textbf{Main Impact} \\
\hline
Jamming &
Communication payload; TT\&C links (HAPS--HAPS, HAPS--UAV, HAPS--satellite) &
Loss of availability; degraded SNR; link outages \\
\hline
Replay &
TT\&C; navigation and telemetry paths &
False state estimation; misguidance; control confusion \\
\hline
System Intrusion &
C\&DH; onboard computer system; ground segment interfaces &
Unauthorized control; data exfiltration; malware persistence \\
\hline
Data Manipulation \& Identity Deception &
Telemetry sensors; routing/control traffic; GPS receivers &
Integrity loss; spoofed identities; unsafe decisions \\
\hline
Adversarial ML Attacks &
AI-based modulation, channel estimation, resource allocation, perception &
Misclassification; suboptimal resource use; degraded autonomy \\
\hline
DoS/DDoS &
Communication payload; TT\&C; gateway and NOC interfaces &
Bandwidth/CPU exhaustion; command-channel starvation \\
\hline
Supply Chain (Software/Hardware) &
All subsystems (COTS components, firmware, toolchains) &
Stealthy backdoors; Trojanized components; systemic compromise \\
\hline
\end{tabular}%
}
\end{table}

\subsection{Key Attack Categories}
We adopt the following terminology for the attack types examined in this work:
\begin{itemize}
    \item \textbf{Jamming}: Deliberate interference within victim frequency bands, degrading signal-to-noise ratios and disrupting service availability.
    \item \textbf{Replay}: Capturing and retransmitting legitimate signals or data packets to deceive receivers, commonly targeting GPS and telemetry systems.
    \item \textbf{System Intrusion}: Gaining unauthorized system access via software vulnerabilities, malware deployment, or credential compromise.
    \item \textbf{Data Integrity and Identity Deception}: Combined threats encompassing (i) data manipulation attacks that modify transmitted information while maintaining sender authenticity, and (ii) identity spoofing attacks that impersonate legitimate network entities to establish unauthorized channels.
    \item \textbf{Adversarial Attacks}: Carefully crafted input perturbations designed to cause machine learning models to misclassify data or make suboptimal resource allocation decisions in AI-driven HAPS systems.
    \item \textbf{DoS/DDoS}: Resource exhaustion attacks targeting bandwidth or computational capacity through protocol exploitation, traffic flooding, or amplification techniques.
    \item \textbf{Supply Chain}: Inserting malicious code or hardware components during system development, integration, or distribution phases, including both software and hardware supply chain vulnerabilities.
\end{itemize}

\subsection{Jamming Attacks}
\begin{figure}[ht!]
    \centering
    \small
    \begin{tikzpicture}[
        node distance=6mm and 10mm,
        act/.style={rounded corners, draw, align=center, inner sep=3pt},
        arrow/.style={-{Latex[length=2mm]}}
    ]
        \node[circle,fill,inner sep=1.5pt] (start) {};
        \node[act, below=of start] (a1) {Attacker configures jammer\\(barrage/reactive)};
        \node[act, below=of a1] (a2) {Emit interference\\(wideband noise / tones)};
        \node[act, below=of a2] (a3) {Victim link SNR drops\\(pilot/sync corrupted)};
        \node[act, below=of a3] (a4) {Packet errors, re-acquisition,\\service outage in footprint};
        \node[draw, circle, inner sep=3pt, below=8mm of a4] (end) {};
        \node[circle, fill, inner sep=1.5pt] at (end) {};
        \draw[arrow] (start) -- (a1);
        \draw[arrow] (a1) -- (a2);
        \draw[arrow] (a2) -- (a3);
        \draw[arrow] (a3) -- (a4);
        \draw[arrow] (a4) -- (end);
    \end{tikzpicture}
    \caption{Jamming attack activity: interference degrades SNR and disrupts link acquisition.}
    \label{jamming}
\end{figure}
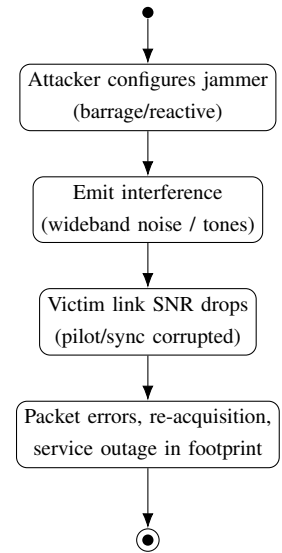

Jamming denotes intentional interference within victim bands that degrades signal-to-noise ratio (SNR) and disrupts reliable reception. Adversaries may use barrage or sweeping noise jammers against HAPS–UAV control links (e.g., 2.4/5 GHz) \cite{ieee_802_11_1997}, or reactive/spot jammers targeting pilot or synchronization symbols. Wideband noise or high-power tones near the carrier collapse the link budget, increasing error rates and causing loss of acquisition; persistent interference can yield outages over the affected footprint. Figure \ref{jamming} illustrates a representative scenario.

Note that identity deception (e.g., GPS or node impersonation) is not a jamming phenomenon and is treated separately in Subsection ``Identity Spoofing Attacks.''

\subsection{Replay Attacks}
\begin{figure}[ht!] 
    \centering
    \includegraphics[width=.4\textwidth]{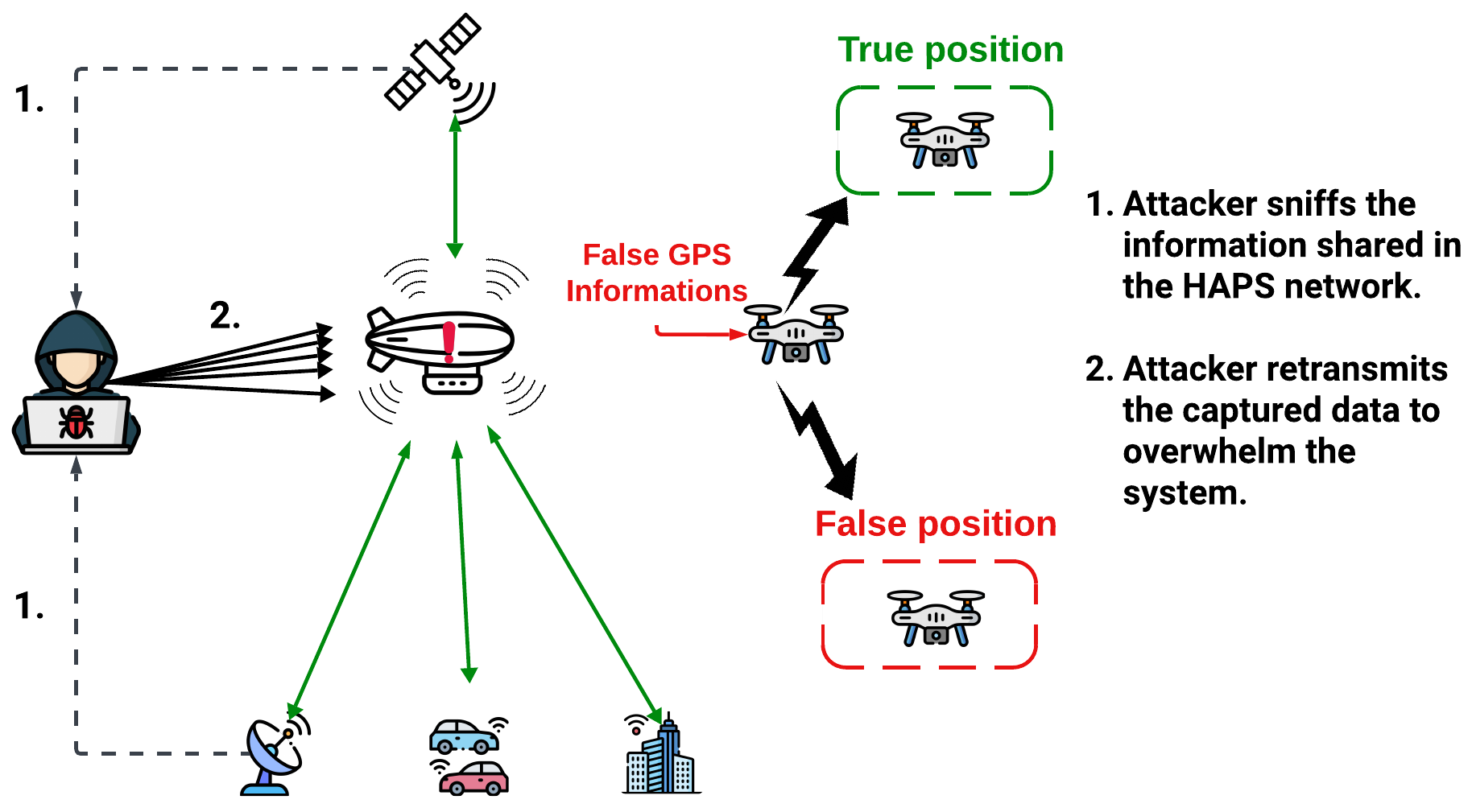}
    \caption{Replay attacks}
    \label{fig:replay}
\end{figure}

Replay refers to the capture and retransmission of valid signals or data to mislead receivers or control logic. Practical realizations include SDR-based capture and rebroadcast of command, telemetry, or GPS signals with adjusted timing, which can induce incorrect state estimation or position fixes, as depicted in Figure \ref{fig:replay}. Similar tactics apply to video feeds by substituting prerecorded frames for live streams.

\subsection{System Intrusion Attacks}
\begin{figure}[ht!] 
    \centering
    \includegraphics[width=0.5\textwidth]{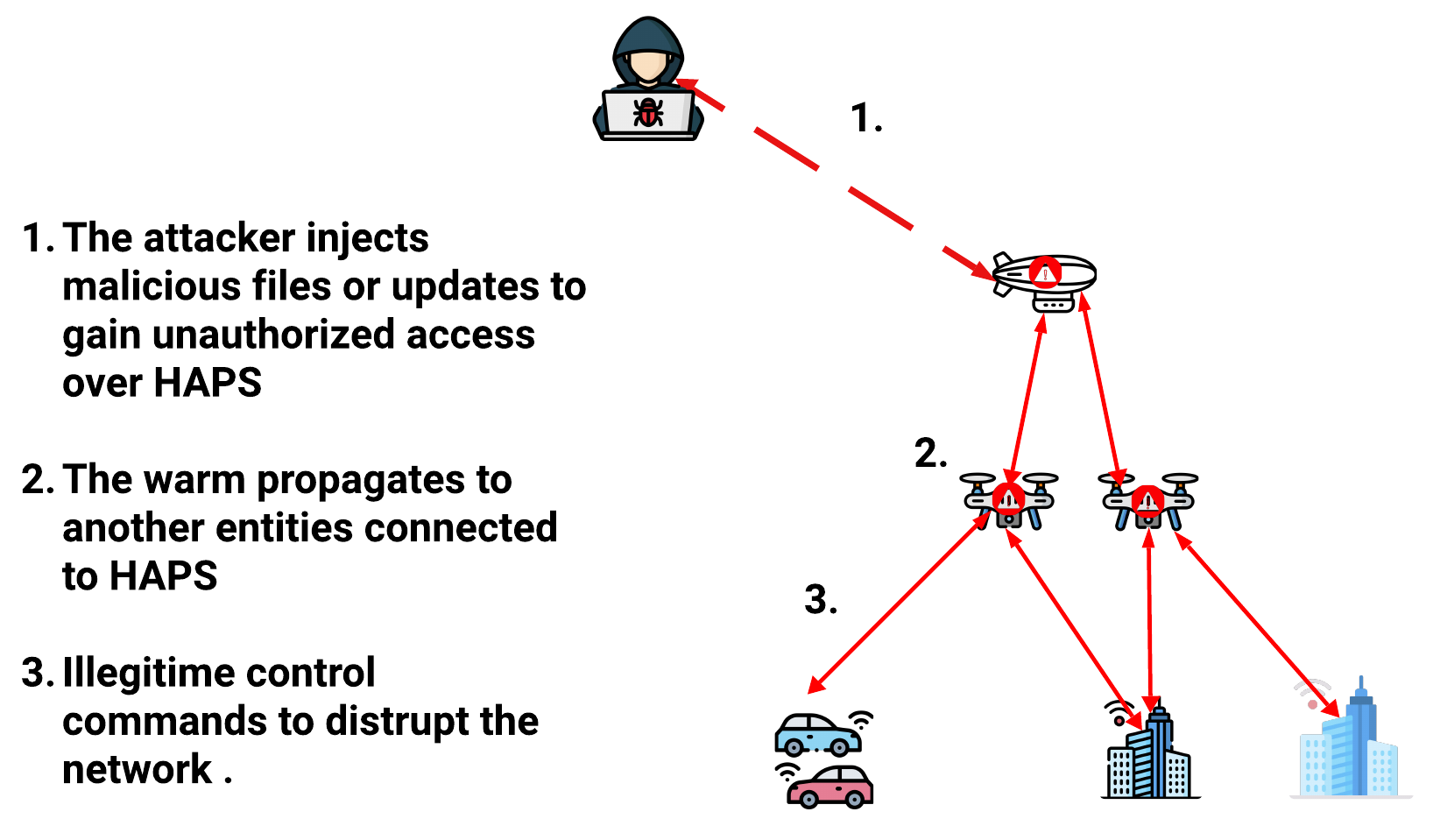}
    \caption{System intrusion attacks}
    \label{fig:intrusion}
\end{figure}

\begin{sloppypar}
System intrusion occurs when attackers illegally exploit vulnerabilities in the HAPS system to gain unauthorized access or make unauthorized modifications to systems, resources, or data to take control, as shown in Figure \ref{fig:intrusion}. The typical methods used for system intrusion attacks involve exploiting system weaknesses, viruses, Trojans, worms, and other harmful software. HAPS can be illegally invaded or remotely controlled by exploiting vulnerabilities or buffer overflows in the system. Trojan attacks consist of unauthorized access to the HAPS system through the utilization of Trojan horses, distributed hacker Trojan horses, and persistent Trojan horses. System breaches frequently result in system harm, including incidents like data exposure, file removal, and forced password decryption, which can endanger the security of both individuals and organizations.
\end{sloppypar}

A worm attack is a form of network attack that relies primarily on the damaging capabilities of viruses to disrupt user data or networks, carry out planned destruction, and initiate remote attacks. Viruses can manipulate the HAPS network. Similar to UAV sensor attacks discussed in security surveys \cite{shafique2021security}, the flight control subsystem of HAPS can also be affected. Injection of fault data serves as a type of virus. It may result in an inability to control altitude.

\subsection{Data Integrity and Identity Deception Attacks}
We explicitly distinguish between two classes of threats: (i) \textit{Data Manipulation Attacks}, which target information integrity while the sender identity remains authentic, and (ii) \textit{Identity Spoofing Attacks}, which impersonate legitimate entities to subvert trust relationships regardless of the specific payload content.

FANETs' decentralized nature and open wireless medium expose them to two primary categories of deceptive attacks: data manipulation attacks that compromise information integrity, and identity spoofing attacks that exploit trust relationships through impersonation.
\subsubsection{Data Manipulation Attacks}
Data manipulation attacks target the integrity of transmitted information without necessarily compromising the sender's identity. In HAPS-coordinated FANETs, attackers exploit data link vulnerabilities to alter critical flight parameters such as GPS coordinates, velocity vectors, or mission instructions while maintaining the appearance of legitimate communication sources.

These attacks can manifest through:
\begin{itemize}
    \item \textbf{In-transit Data Alteration}: Intercepting and modifying data packets during wireless transmission between HAPS and UAV nodes
    \item \textbf{Sensor Data Corruption}: Targeting individual sensor readings to cause gradual deviation from intended flight paths
    \item \textbf{Command Injection}: Inserting malicious commands into legitimate control sequences
\end{itemize}

The primary objective is to cause operational disruption through corrupted decision-making processes while avoiding immediate detection by maintaining communication authenticity.

\subsubsection{Identity Spoofing Attacks}
Identity spoofing attacks focus on impersonating legitimate network entities to gain unauthorized access or establish malicious control channels. Unlike data manipulation attacks, these attacks primarily target authentication and trust mechanisms rather than data content integrity.

\begin{figure}[ht!]
    \centering
    \includegraphics[width=.4\textwidth]{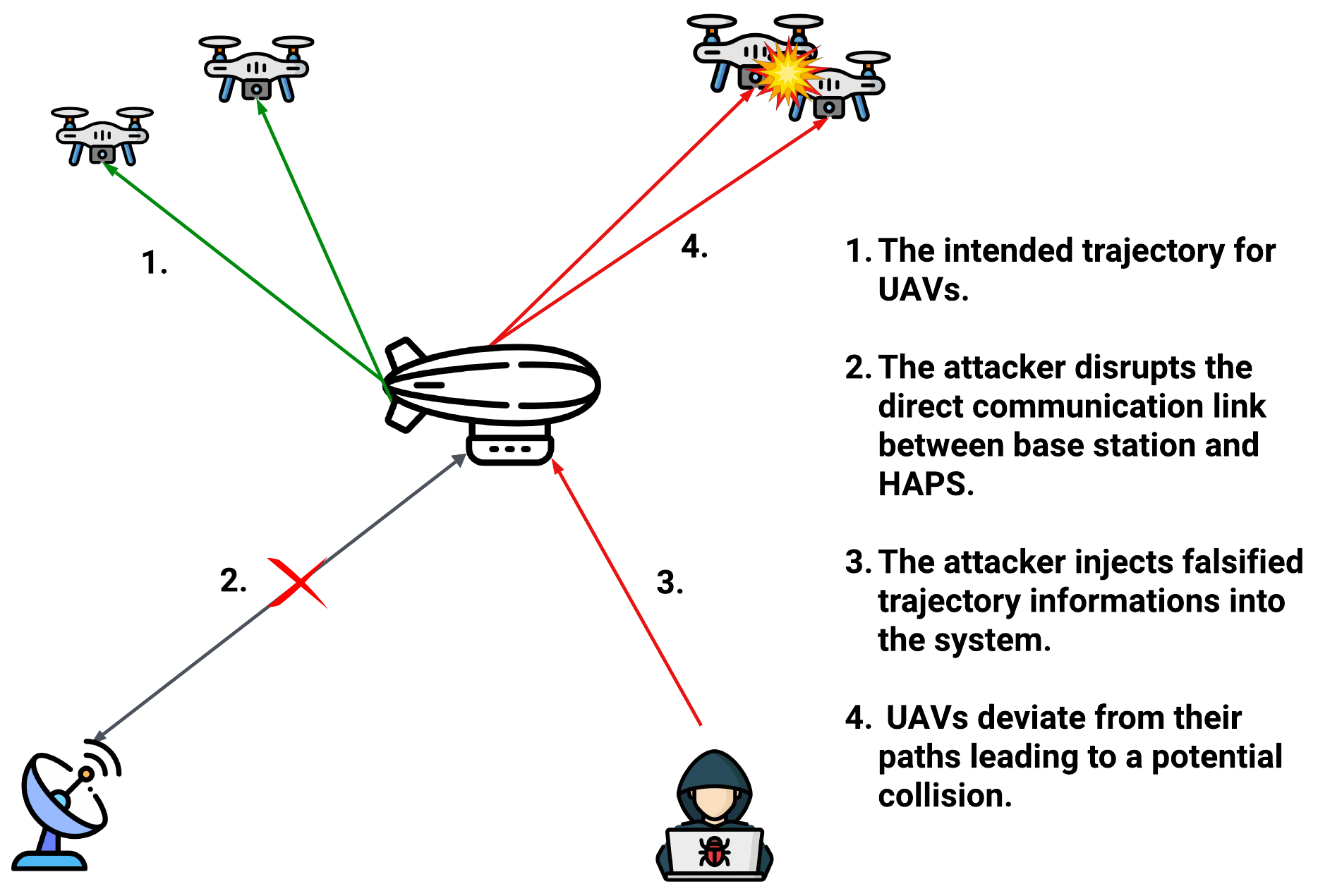}
    \caption{Identity Spoofing Attack (GPS Spoofing): Malicious node impersonating legitimate HAPS to deceive UAV navigation}
    \label{fig:spoofing}
\end{figure}

Key spoofing attack vectors include:
\begin{itemize}
    \item \textbf{Node Impersonation}: Attackers masquerade as legitimate HAPS or UAV nodes to inject false routing information or establish unauthorized communication channels
    \item \textbf{GPS Spoofing}: Broadcasting false GPS signals to deceive navigation systems, as illustrated in Figure \ref{fig:spoofing}
    \item \textbf{Authentication Bypass}: Exploiting vulnerabilities in routing protocols (e.g., AODV) to establish illegitimate network presence
\end{itemize}

The decentralized nature of FANETs makes detection and mitigation of these attacks particularly challenging, as there is no centralized authority to verify data integrity or authenticate node identities consistently across the network.

\subsection{Adversarial Attacks}

\begin{figure}[ht!]
    \centering
    \includegraphics[width=0.5\textwidth]{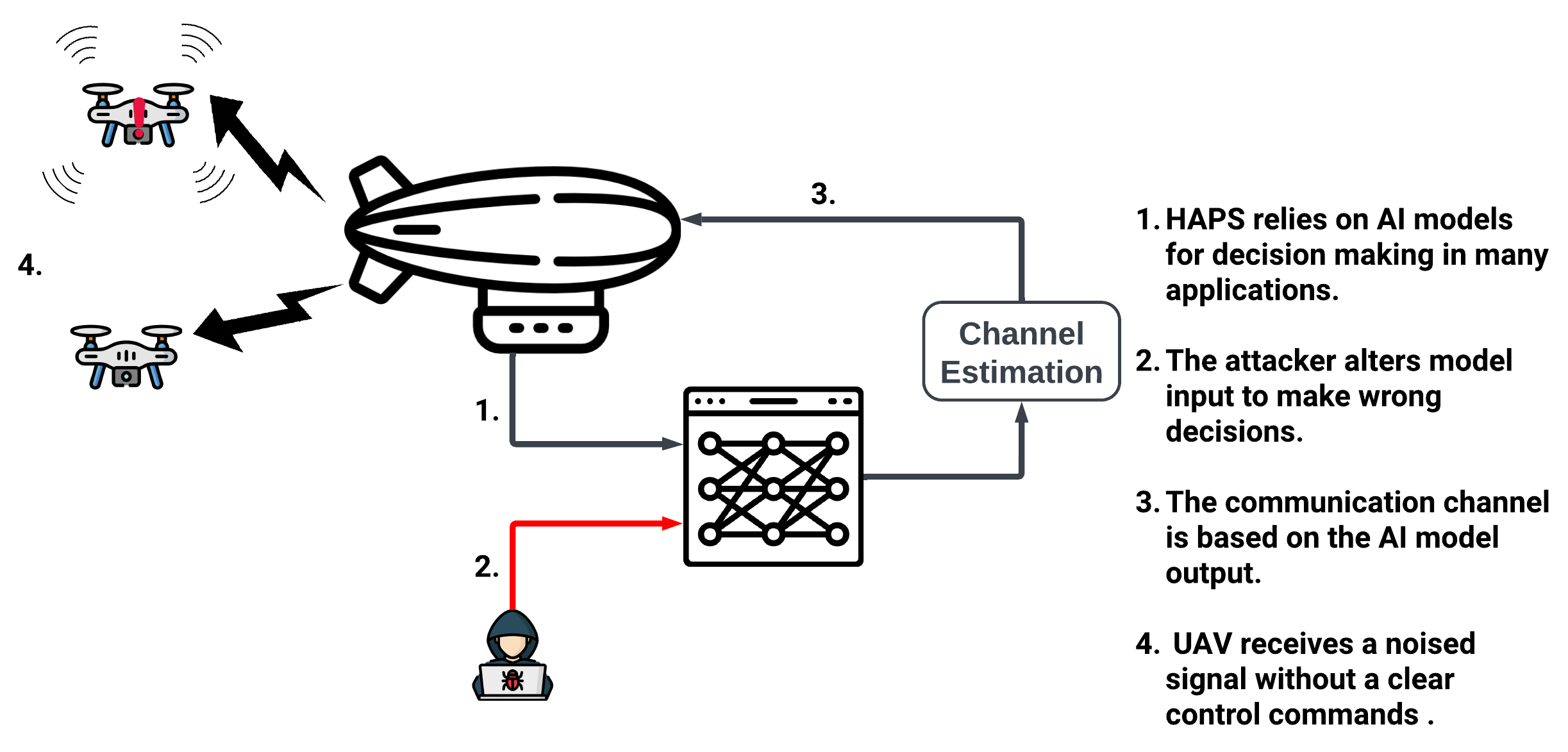}
    \caption{Adversarial Attacks on AI-based signal modulation in HAPS system}
    \label{fig:adversarial}
\end{figure}

HAPS systems increasingly rely on AI models for critical functions such as radio signal modulation, channel estimation, resource allocation, beamforming optimization, and interference mitigation. The functionality of these learning-based models can be severely compromised by adversarial attacks, sophisticated techniques that deliberately manipulate input data to cause ML or DL models to generate incorrect predictions or classifications, as illustrated in Figure \ref{fig:adversarial}.

\subsubsection{Adversarial Attack Mechanisms in HAPS}

Adversarial attacks against HAPS systems exploit the inherent vulnerabilities of AI models through carefully crafted perturbations that are often imperceptible to human operators but catastrophic to automated systems. These attacks can manifest through several vectors:

\textbf{Signal Modulation Attacks}: Adversarial perturbations can be injected into communication signals to deceive DL-based modulation recognition systems. Prior work \cite{zhang2019adversarial} demonstrated that adding minimal noise perturbations to communication signals can significantly reduce the accuracy of CNN-based modulation classifiers. In HAPS contexts, this results in communication failures between high-altitude platforms and ground stations, potentially disrupting entire communication networks.

\textbf{Information Retrieval Attacks}: HAPS platforms increasingly rely on deep product quantization networks for efficient data storage and retrieval in distributed systems. Chen et al. \cite{chen2023adversarial} demonstrated how adversarial examples can be generated to mislead deep product quantization networks used in image retrieval systems, with implications for HAPS data management systems that employ similar quantization techniques for efficient storage and retrieval of sensor data and communication metadata.

\textbf{Channel Estimation Attacks}: HAPS systems employ neural networks for channel state information (CSI) estimation to optimize beamforming and resource allocation. Adversarial attacks targeting these systems can manipulate channel estimation, leading to suboptimal resource allocation and degraded network performance \cite{zhang2019adversarial}.

\textbf{Power Allocation Attacks}: Sahay et al. \cite{defending} analyzed adversarial attacks on deep learning-based power allocation in Massive MIMO networks. Their results show that carefully crafted perturbations to the network inputs can induce significant utility loss and QoS violations by driving suboptimal power allocation decisions.

Usama et al. \cite{adversarial_jamming} analyzed wireless network vulnerabilities and demonstrated that carefully crafted adversarial examples can evade intrusion detection systems while maintaining attack effectiveness. These findings indicate that ML-based defenses, if not stress-tested against adversarial inputs, may provide a false sense of security and can be circumvented with minimal additional resources.

Shi et al. \cite{shi2021generative} demonstrated adversarial attacks against automatic modulation classification using generative adversarial networks, while Sagduyu et al. \cite{LEB2023} demonstrated adversarial deep learning attacks for over-the-air spectrum poisoning.

\subsubsection{Attack Sophistication and Countermeasures}

Modern adversarial attacks against HAPS systems demonstrate increasing sophistication:

\textbf{Universal Adversarial Perturbations}: These attacks generate perturbations that remain effective across different input samples and network conditions. Experimental results in prior work \cite{zhang2019adversarial,kokalj2018adversarial} show that universal perturbations crafted for CNN-based automatic modulation classification can achieve high attack success rates across different modulation schemes, demonstrating transferability that makes them effective against diverse HAPS deployments.

\textbf{Physical Layer Attacks}: Unlike digital perturbations, physical adversarial attacks manipulate the actual transmitted signals. Kokalj-Filipovic et al. \cite{kokalj2018adversarial} demonstrated adversarial examples for the physical world, showing how machine learning systems in the RF domain can be attacked. Their over-the-air experiments showed that these attacks can maintain effectiveness even after passing through realistic channel conditions, including atmospheric effects typical in HAPS communications.

\textbf{Advanced Attack Techniques}: Beyond RF tasks, perception-driven autonomy is also vulnerable: shape-sensitive adversarial patches can comprehensively disrupt monocular depth estimation pipelines used for navigation \cite{guesmi2024_iros_ssap}. Additionally, training-time data poisoning has been demonstrated via incremental physical adversarial attacks that subvert model learning \cite{alqudah2023_icc_graybox}.

\subsection{DoS/DDoS Attacks}
DDoS attacks overwhelm HAPS platforms by saturating control and payload links with excessive traffic from distributed sources, depleting bandwidth, CPU, and memory, as illustrated in Figure \ref{fig:ddos}. In stratospheric settings, long link budgets and asymmetric backhaul can complicate rate limiting and traceback, while intermittent connectivity hinders timely key refresh and blacklist propagation. Common vectors include reflection/amplification (e.g., DNS, NTP, SSDP), protocol abuse (e.g., TCP SYN floods, QUIC/HTTP/2 concurrency abuse), ECN/DSCP (DiffServ Code Point) manipulation that degrades QoS, and application-layer floods that target TT\&C endpoints. These threats are acute for HAPS given the limited on-board compute and the criticality of command and control channels.

\begin{itemize}
        \item Bandwidth/resource consumption disrupting communication and control systems
        \item Protocol vulnerability exploitation targeting network layer weaknesses
        \item Memory exhaustion through corrupted data packet floods
\end{itemize} 

\begin{figure}[ht!]
    \centering
    \small
    \begin{tikzpicture}[
        node distance=6mm and 10mm,
        act/.style={rounded corners, draw, align=center, inner sep=3pt},
        arrow/.style={-{Latex[length=2mm]}}
    ]
        \node[circle, fill, inner sep=1.5pt] (start) {};
        \node[act, below=of start] (d1) {Forge source IP to reflector\\(amplification attack)};
        \node[act, below=of d1] (d2) {Reflector sends amplified response\\to victim HAPS};
        \node[act, below=of d2] (d3) {Bandwidth exhaustion / queue overflow\\at HAPS};
        \node[act, below=of d3] (d4) {HAPS resources depleted};
        \node[draw, circle, inner sep=3pt, below=8mm of d4] (end) {};
        \node[circle, fill, inner sep=1.5pt] at (end) {};
        \draw[arrow] (start) -- (d1);
        \draw[arrow] (d1) -- (d2);
        \draw[arrow] (d2) -- (d3);
        \draw[arrow] (d3) -- (d4);
        \draw[arrow] (d4) -- (end);
    \end{tikzpicture}
    \caption{DDoS amplification activity: forged requests to reflectors generate amplified replies toward HAPS.}
    \label{fig:ddos_amp_flow}
\end{figure}
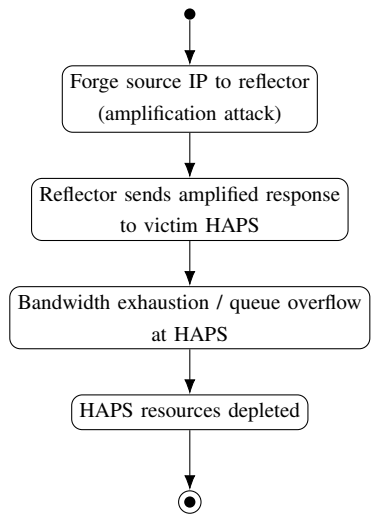

\begin{figure}[ht!] 
    \centering
    \includegraphics[width=.4\textwidth]{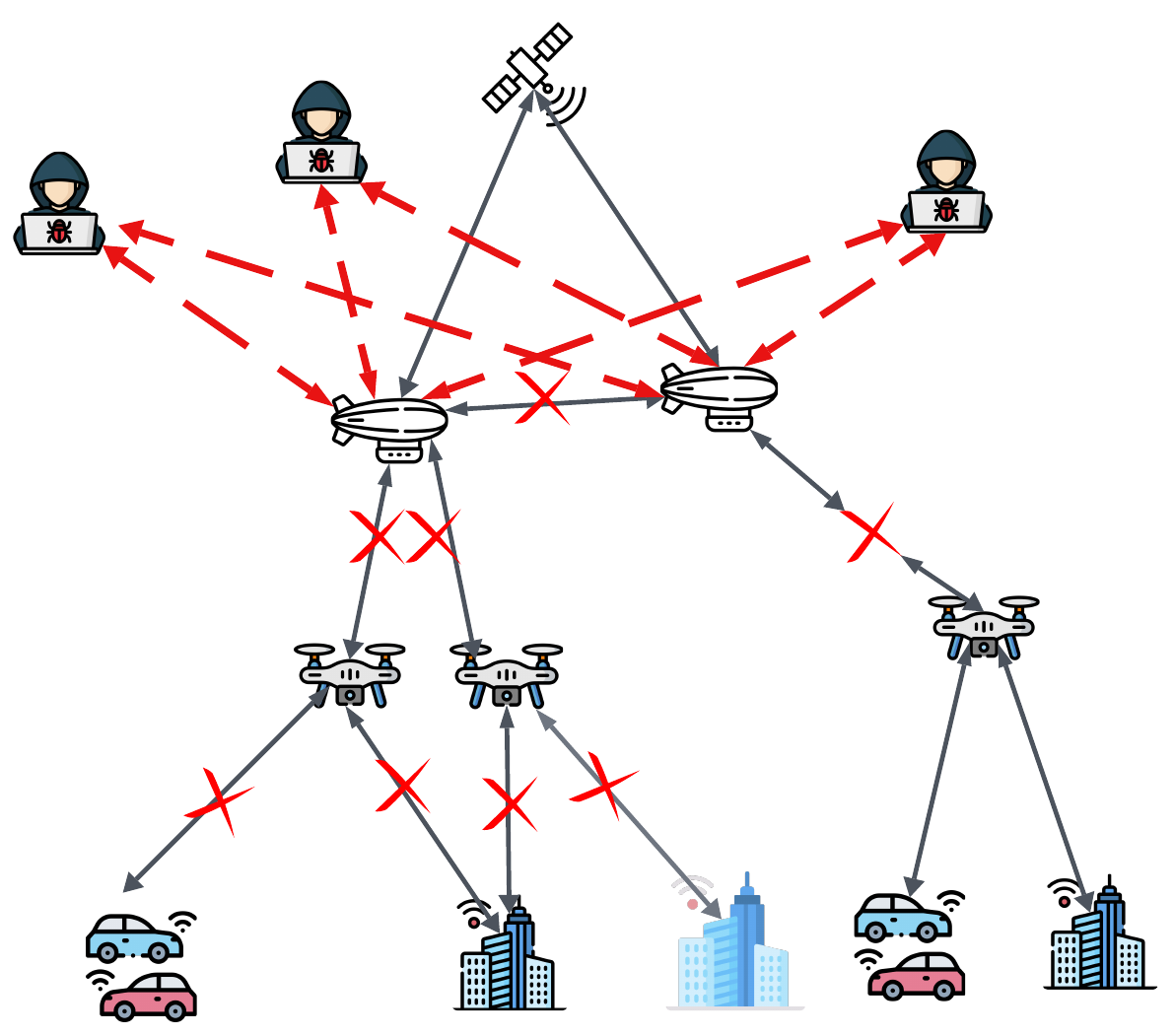}
    \caption{DDoS Attack: Overwhelming HAPS by floods of traffic data to disrupt legitimate communication links}
    \label{fig:ddos}
\end{figure}

	\textbf{Attack Types:} Bandwidth depletion attacks include UDP floods, fraggle attacks using reflectors, and amplification attacks targeting protocol vulnerabilities, as shown in Figure \ref{fig:ddos_amp_flow}. ToS floods manipulate ECN/DSCP flags to disrupt quality of service mechanisms. Resource depletion attacks exploit TCP SYN floods, creating half-open connections, ping floods from distributed sources, and malformed packets (land attacks, ping of death), causing buffer overflows \cite{salim}.

\subsection{Supply Chain Attacks}
Supply-chain attacks exploit trusted software and hardware dependencies to compromise systems during development, integration, or distribution. For HAPS, such attacks may target toolchains, code-signing infrastructure, firmware update paths, or component provenance, and can yield broad compromise due to transitive trust. Attackers commonly abuse third-party relationships and weak controls to implant malware, exfiltrate secrets, or introduce modified components that evade routine testing.    

Recently, Nobelium, a Russian threat actor, has been targeting organizations considered crucial to the worldwide IT supply chain. Microsoft states that resellers are the target of these attacks. Attackers have not tried to take advantage of software flaws, but instead have employed tactics such as phishing to steal valid credentials, obtain privileged access, and use this to infiltrate the system \cite{ico}.

\begin{figure}[ht!] 
            \centering
            \includegraphics[width=.4\textwidth]{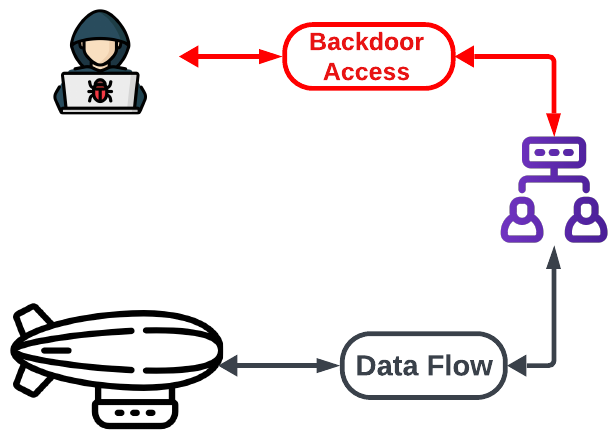}
            \caption{The supply chain attacks}
            \label{fig:supplychain}
\end{figure}

	\textbf{Software Supply Chain Attacks}: 
Due to the increasing interdependence of security within supply chains, attackers are now more interested in carrying out supply chain attacks \cite{ico}, making the HAPS system a vulnerable target for such assaults. A supply chain attack occurs when multiple attacks are combined to breach the information security of suppliers and their customers. In this situation, HAPS and the software suppliers integrated into these platforms are likely targets for these attacks. The initial strike is conducted in order to obtain the assets of a HAPS software provider, which are then utilized to target either the HAPS system itself or another provider while developing these aircrafts. Supply chain attacks exploit the security reliance of a connected network of suppliers, operators, and users of HAPS software and components. Typically, these attacks are intricate and require months of planning before being carried out, often remaining unnoticed for an extended period of time. The assailants frequently concentrate on embedding backdoor entry in HAPS suppliers' code and software to carry out additional attacks on these platforms, utilizing malware and HAPS operators' trust in suppliers as main tactics, as shown in Figure \ref{fig:supplychain}. Attackers can take advantage of weaknesses in the supply chain to breach data from one part of the HAPS design chain, affecting various subsystems or components through compromised supplier systems.

In practice, reliance on commercial off-the-shelf (COTS) components and third-party firmware further enlarges the attack surface. While COTS accelerates deployment and reduces cost, opaque provenance and variable patch cadences complicate vulnerability management and increase exposure to compromised updates and counterfeit parts.

\textbf{Hardware Supply Chain Attacks}: Hardware attacks in the supply chain process present additional risks to HAPS, particularly through modified or malicious components embedded in HAPS subsystems. Attacks may involve insider threats, malicious hardware with hidden functionalities, or malicious firmware embedded in hardware parts, stored in persistent memory without integrity protection. Additionally, Trojan components, mimicking genuine parts, could be inserted during fabrication, making HAPS systems vulnerable to undetectable attacks. This part examines the commonly known supply-chain attacks on  Integrated Circuits (IC) components\cite{bhunia2016hardware}:

\begin{enumerate}
    \item \textbf{Insider Threat}: The risk of this attack is an ongoing worry that exists at all points in the supply-chain process. At any given moment, an individual on the inside with access has the potential to modify or include features, steal confidential data or design files, introduce or swap components, or conduct different attacks that could put the security of the entire product at risk.
    \item \textbf{Malicious Hardware}: Malicious hardware refers to a component that fulfills necessary functions and meets design requirements but includes additional or hidden functionalities that execute malicious actions continuously or upon specific activation inputs. Due to the vast number of potential input vectors, exhaustive testing of such components is generally impractical if not statistically impossible in most cases.
    \item \textbf{Malicious Firmware}: Many hardware parts in HAPS system need embedded software, commonly referred to as firmware, to function correctly. Just like malicious hardware, firmware could have hidden or extra features. Additionally, a notable portion of this software is stored in persistent memory without any integrity protection or authentication before running. As a result, the firmware may be intentionally harmful from the beginning or altered by different individuals throughout the supply chain process.
    \item \textbf{Insertion of Trojan Component}: A Trojan component is a part designed to be functionally equivalent to a genuine device but includes covert logic that activates under rare or context-specific conditions. Evasion mechanisms include: (i) \textit{Trigger rarity} (e.g., long counters, magic input patterns, voltage/temperature windows) so typical manufacturing tests never activate the payload; (ii) \textit{Low switching activity} and \textit{clock/power gating} so dynamic power signatures remain within process variation; (iii) \textit{Analog/RF side-channels} (e.g., slight biasing changes) that elude purely digital test vectors; (iv) \textit{Camouflaged gates/layout obfuscation} that defeat optical imaging and reverse engineering; and (v) \textit{Firmware-assisted Trojans} where malicious microcode is stored in non-volatile memory without integrity checks. 
\end{enumerate}

\section{Regulatory and Standardization Landscape}
Deploying secure HAPS infrastructure requires navigating complex regulatory and standards frameworks governing NTNs. Technical solutions must align with evolving policy requirements and international coordination mechanisms.

\subsection{Spectrum, Service, and Safety Coordination}

\textbf{ITU Coordination:} HAPS spectrum operations fall under ITU Radio Regulations, requiring coordination across service allocations. Security implementations must consider encryption export controls, interference management protocols, and lawful access requirements mandated by member administrations \cite{itu_f1500_2000,5,6}.
    
\textbf{Aviation Safety Integration:} ICAO guidance on HAPS airworthiness and flight safety directly impacts cybersecurity through requirements for secure command/control links, robust detect-and-avoid systems, and reliable CNS integrity. National aviation authorities (EASA, FAA) impose additional risk management and incident reporting obligations that influence security architecture design.

\subsection{Telecom Standards for NTN Security}
    	\textbf{3GPP (5G/6G with NTN features):} 3GPP work groups for NTNs extend identity management, authentication, and key management to space/airborne segments \cite{13,asad2025_jsac_ztfl}. For HAPS acting as part of 3GPP access/backhaul, relevant topics include SUPI/Subscription security, NAS/RRC security over stratospheric channels, interworking with satellite components, and lawful intercept compliance. Practical implications include cipher suite selection for high-latency links, key refresh policies under intermittent connectivity, and secure handover across NTN/terrestrial domains.

\subsection{Cybersecurity Baselines and Certification}
    	\textbf{NIST and ISO/IEC:} Baselines such as NIST SP 800-series and ISO/IEC 27001/27002 provide controls for access management, crypto, logging, and incident response \cite{nist_sp_800_53_r5_2020,iso_iec_27001_2022,iso_iec_27002_2022}. For avionics software/hardware, secure development and SBOM practices (e.g., NIST SSDF) strengthen supply-chain assurance \cite{nist_sp_800_218_2022}. For the software weakness taxonomy, we reference CWE enumerations to categorize and track common weakness classes across the codebase.
    
        \textbf{ETSI and ENISA:} ETSI standards for cybersecurity (e.g., ETSI EN 303 645 for connected devices, ETSI TS 103 457 for threat information sharing) and ENISA guidance on space/aviation security inform vulnerability disclosure, telemetry, and cross-stakeholder incident coordination for HAPS ecosystems \cite{etsi_en_303_645_2020,etsi_ts_103_457_2023,enisa_space_threat_landscape_2023}.
    
        \textbf{DO-326A/ED-202A family (Airworthiness Security):} RTCA/Eurocae guidelines (aircraft system security process, information security framework) are instructive for HAPS with aviation-adjacent safety requirements, supporting security by design, risk assessment, and continued airworthiness \cite{rtca_do_326a_2014,eurocae_ed_202a_2014}.

\subsection{Compliance Implications for HAPS}
From a deployment perspective, HAPS operators should:
\begin{itemize}
    \item Implement spectrum-compliant encryption and interference management aligned with ITU coordination outcomes \cite{itu_f1500_2000,5}.
    \item Adopt 3GPP NTN security profiles for identity, authentication, and key management when interoperating with mobile networks \cite{13}.
    \item Apply NIST/ISO baselines for configuration, logging, incident response, and SBOM-based supply-chain assurance across flight and ground segments.
    \item Prepare for aviation-grade security certification where applicable (e.g., using DO-326A/ED-202A processes) and align with ICAO safety objectives.
    \item Establish cross-border incident reporting and threat-intel sharing procedures per ETSI/ENISA guidance.
\end{itemize}

These frameworks contextualize the technical defenses proposed in this manuscript and highlight the importance of harmonizing link-layer security, software assurance, and operations with policy and certification pathways for real-world HAPS deployments.

\section{Defense Mechanisms}

\begin{figure}[htbp]
    \centering
   \includegraphics[width=.5\textwidth]{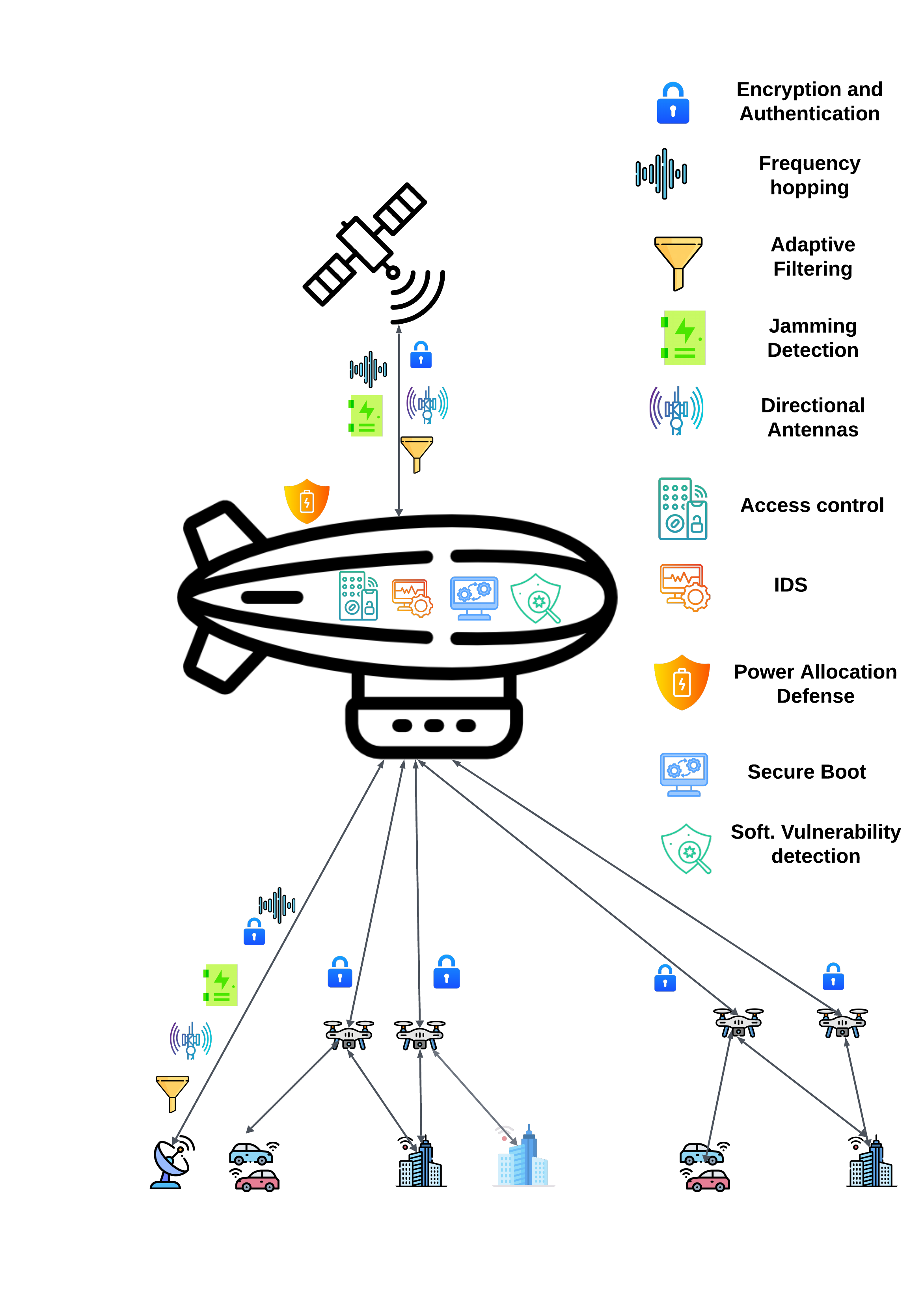}
    \caption{Cybersecurity defenses against HAPS attacks}
    \label{fig:defense}
\end{figure}

\begin{sloppypar}
To secure the HAPS platform and mitigate potential threats, various defense mechanisms need to be applied across different subsystems, including communication, control, and power management. These mechanisms ensure the integrity of data transmission, prevent unauthorized access, and enhance the overall system resilience. The proposed mitigations should be distributed across HAPS components, such as the onboard embedded systems, ground stations, and communication links between them, as shown in Figure \ref{fig:defense}. \textit{Encryption and authentication} should be integrated into the \textit{communication subsystem} to protect data transmission between HAPS, ground stations, and satellites. \textit{Frequency hopping} and \textit{directional antennas} further safeguard communications by preventing jamming attacks and ensuring focused signal transmission. In the \textit{control subsystem}, \textit{IDS} and \textit{access control} mechanisms monitor and restrict unauthorized access. The \textit{power management subsystem} benefits from a \textit{power allocation defense mechanism} to protect against energy distribution attacks. Additionally, \textit{secure boot} and \textit{software vulnerability detection} are critical across all subsystems to guarantee system integrity. These defense mechanisms are detailed below.
\end{sloppypar}
\subsection{Encryption and Authentication}
In \cite{haps}, Lou et al. mentioned that HAPS is required to establish communication links with all the different parts of its network, and with these connections, global information will be shared. The communication that occurs through the network presents reliability and security issues, so implementing strong encryption protocols ensures secure data transmission and protects against interception and unauthorized access. Using strong encryption and authentication protocols can protect HAPS system from multiple cyber threats. For example, in the case of spoofing attacks, by using an authentication mechanism only the legitimate signals are processed.

\subsection{Frequency Hopping}
Employing frequency hopping spread spectrum (FHSS) techniques can make it difficult for jammers to target specific frequencies, as the communication frequency changes rapidly and unpredictably. FHSS techniques involve repeatedly switching the carrier frequency according to a pseudorandom hop sequence shared between the transmitter and the
receiver to decrease interference and prevent unauthorized interception. The deployment of FHSS in stratospheric environments presents unique challenges requiring specialized solutions that differ significantly from terrestrial implementations:

	\textbf{Stratospheric Implementation Challenges:} HAPS platforms experience complex velocity profiles with horizontal velocities of 50--80~m/s and vertical velocities up to 10~m/s \cite{abalos_2022}. For Ka-band frequencies, Doppler effects create time-varying profiles requiring compensation across ground links (3--8~kHz shifts), HAPS-to-HAPS links (6--15~kHz), and HAPS-satellite links (20--45~kHz). Stratospheric turbulence introduces rapid fluctuations with standard deviations of 2--8~m/s, requiring Kalman filtering for real-time prediction within $\pm 500$~Hz accuracy.

	\textbf{Propagation and Hardware Constraints:} Specific molecular absorption in the stratospheric environment requires careful frequency planning \cite{itu_p676_2022}. Temperature inversions create multipath propagation with delay spreads of 5--25~$\mu$s. Commercial FHSS commonly operates at 2.5~hops/s, inadequate for anti-jamming, while military systems achieve 10{,}000--50{,}000~hops/s \cite{torrieri_2015}. SDR platforms face switching limitations of 50--200~$\mu$s, constraining hop rates to 1--5~kHz \cite{reed_2019}. HAPS-optimized implementations require custom FPGA synthesizers achieving sub-10~$\mu$s switching for effective anti-jamming rates exceeding 10~kHz.

	\textbf{Power and Synchronization Considerations:} High-speed switching increases power consumption by 40--60\%, critical for solar-powered platforms. GPS-disciplined oscillators provide $10^{-12}$ stability \cite{trimble_2018}, but stratospheric deployment faces 2--4~dB signal reduction and variable path delays requiring advanced compensation algorithms.

\textbf{Multi-HAPS Constellation Synchronization:} Distributed HAPS constellations rely on Byzantine fault-tolerant consensus mechanisms to sustain coordinated frequency-hopping patterns even under intermittent or degraded GPS conditions. In GPS-denied periods, inertial navigation systems coupled with high-stability cesium or rubidium atomic clocks provide holdover performance on the order of $10^{-11}$ over extended durations, ensuring the continuity of timing references across the fleet. The Inter-HAPS range using ultra-wideband signals further refines the relative alignment of the clock, achieving high-precision synchronization that is suitable for coordinated spectrum management and secure frequency-hopping operations \cite{PEREZSOLANO202444}.

\textbf{Advanced Cognitive Radio Spectrum Coordination:} HAPS frequency-hopping systems must coexist with satellite services in shared spectrum bands while maintaining strong anti-jamming performance. Real-time spectrum sensing, combining energy detection with cyclostationary feature extraction, enables reliable identification of primary satellite users. Machine-learning approaches—particularly Long Short-Term Memory (LSTM) architectures—enhance spectrum awareness by learning temporal signal characteristics and improving detection accuracy under low SNR conditions \cite{s22062286}. These models can be used to predict satellite activity patterns and adapt FHSS sequences to maintain isolation from primary satellite services while preserving resilience against intentional interference. Dynamic spectrum databases, updated via satellite backhaul links, provide additional situational awareness and interference avoidance capability during ground-station communication outages.

\textbf{Practical Implementation Challenges and Validation:} Laboratory testing of FHSS prototypes under simulated stratospheric conditions reveals several practical implementation challenges. Temperature cycling between extreme cold and warm conditions induces frequency synthesizer drift, necessitating continuous calibration and temperature-compensation mechanisms to maintain hopping stability. In addition, cosmic radiation at very high altitudes leads to substantially elevated radiation exposure levels compared to sea level, increasing the likelihood of single-event effects in digital signal processing hardware. These effects require the use of radiation-tolerant components and error-correcting memory architectures to ensure reliable operation in a near-space environment \cite{REES2024564}.

\textbf{Deployment Feasibility Assessment:} Economic analysis indicates that stratosphere-optimized FHSS implementations cost significantly more than terrestrial equivalents due to radiation-hardened components and environmental qualification requirements. However, the anti-jamming benefits justify these costs for critical infrastructure applications. Recent studies of the HAPS system demonstrate that commercial component-based communication payloads can operate reliably under stratospheric conditions when supported by appropriate thermal and radiation mitigation strategies, validating the technical feasibility of deploying FHSS-capable platforms in the near-space environment \cite{9448899}.


\subsection{Directional Antennas}
Directional antennas provide an effective mitigation strategy against jamming in wireless communication systems by concentrating radiated power toward the intended receiver rather than dispersing it uniformly, as in omnidirectional antennas. By narrowing the beamwidth, directional antennas significantly reduce the angular region through which interference can couple into the receiver chain, thereby limiting a jammer’s ability to disrupt the link. This spatial selectivity enables higher link robustness, maintains stronger signal quality over extended distances, and reduces the number of required relay hops.

Deploying directional antennas on HAPS platforms, however, introduces several operational and environmental challenges. Continuous platform motion requires precise pointing and stabilization, often achieved using multi-axis gimbals and predictive attitude control. Variations in atmospheric density at stratospheric altitudes can shift antenna resonance, while extreme temperature cycles can distort beam patterns. Mechanical constraints limit the size and mass of reflector antennas, whereas phased arrays—though capable of electronic beam steering for multi-link coordination—impose substantial power-consumption requirements.

Recent studies confirm that highly directional links exhibit significantly greater resilience to jamming by exploiting spatial filtering to suppress interference arriving from angles outside the main lobe of the antenna pattern \cite{shrestha2022jamming}.

\subsection{Adaptive Filtering}
Adaptive filtering techniques play a critical role in enhancing the resilience of HAPS communication links against evolving cyber and jamming threats. By dynamically adjusting filter coefficients in real time, adaptive filters can distinguish legitimate communication signals from malicious interference, even when the jamming or attack patterns change rapidly. These filters continuously monitor communication channels or network traffic, detect anomalies, and suppress harmful signal components or malicious data packets before they degrade system performance. 

For HAPS, adaptive anti-jamming filters can update their parameters based on previously observed interference signatures, enabling the system to respond effectively to sophisticated attack types such as distributed denial-of-service (DDoS) flooding or intelligent barrage jamming. Recent work demonstrates that cascaded, low-complexity adaptive filters can achieve significant interference suppression for dynamic wireless environments, validating their relevance for HAPS platforms operating under unpredictable threat conditions \cite{song_adaptive_antijamming_2024}.

\subsection{DoS/DDoS Detection}
As a mitigation strategy to prevent DDoS attacks in satellite networks, machine learning-based approaches \cite{alashhab2022ddos} can be employed to limit malicious network connections. Specifically, we monitor the network's behavior, observing how many echo requests a network entity sends or receives from a set of IP addresses under normal conditions when the network is not experiencing a DoS attack. This baseline behavior is flagged as normal, and any deviations from it are marked as abnormalities, leading to the non-entertainment of ICMP packets in such scenarios.
It is important to clarify that this behavior monitoring does not involve profiling a user's traffic based on their IP address. Rather, the term “normal” refers to an estimation of the average number of ICMP requests a network entity can handle. This information can be easily obtained from the network, even if the IP address allocation is dynamic across networks. This solution prevents both scenarios of ping flooding by (i) limiting the echo requests sent from a network entity in the Network Operations Center (NOC) to multiple ground stations (GSs), thus avoiding overwhelming the communication link of GSs with replies, and (ii) blocking spoofed network devices from sending ICMP requests to a target network entity in GSs, thereby preventing depletion of its network and computational resources. If a network entity sends too many echo requests to different GSs, the NOC will block both the excessive echo requests and their corresponding replies from the GSs. On the other hand, in the second situation, ICMP requests from various sources will be denied at the NOC, and responses will not be returned to the originating IP addresses.
Blocking ICMP requests from certain IP addresses that may cause DDoS attacks is crucial, but legitimate ICMP traffic from users on the HAPS network will not be affected. There might be times when a hacked network element, like a base station (BS), is connected to genuine nodes. However, even in these scenarios, blocking the compromised network entity from creating ICMP requests does not impact the connected legitimate entities, which have their public IP addresses. Because our proposal focuses specifically on ICMP requests from the corrupted node, it does not affect all ICMP packets passing through a transit node.

\subsection{Intrusion Detection Systems (IDS)}
These systems are crucial for early detection of anomalous activities that may indicate a cyber attack, allowing for timely responses, and recent IoT-scale studies show that deep learning can outperform traditional ML for attack detection and device-type identification \cite{hamidouche2024_iot}. Yin et al. \cite{yin} suggest a deep learning method for identifying intrusions by employing recurrent neural networks (RNN-IDS) and evaluating their effectiveness in both binary and multiclass classification assignments. The findings indicate that the RNN-IDS model is more accurate than traditional machine learning approaches. Chawla and colleagues \cite{chawla} introduced a system for detecting intrusions based on anomalies which utilize RNNs incorporating gated recurrent units (GRUs) as well as stacked convolutional neural networks (CNNs) to identify harmful cyber attacks. By analyzing sequences of system calls made by processes, the system creates a standard behavior reference for a specific system. It detects unusual sequences using a language model trained on typical call sequences from the ADFA dataset of system call traces. The writers show that substituting LSTMs with GRUs yields similar results quicker and that integrating GRUs with stacked CNNs enhances anomaly detection. The system under consideration demonstrates encouraging outcomes when identifying abnormal sequences of system calls within the ADFA dataset. Nevertheless, additional studies are required to assess how well it functions in different datasets and real-life situations, as well as to tackle problems linked to adversarial attacks. Xu et al. \cite{xu} introduced a new IDS for detecting network intrusions, comprising a recurrent neural network with GRU, an MLP, and a softmax module. The system demonstrates top performance with detection rates reaching 99.42\% on the KDD 99 dataset and 99.31\% on the NSL-KDD dataset, all while maintaining low false positive rates. Polat and colleagues \cite{polat} presented a technique to enhance the identification of DDoS attacks in SCADA systems that utilize Software Defined Network (SDN) technology. The writers suggest utilizing an RNN classifier model that combines two deep learning techniques in parallel: LSTM and GRU. The suggested model undergoes training and testing using a dataset from an SDN-based SCADA topology that includes both DDoS attacks and normal network traffic information. The findings indicate that the RNN model suggested has an accuracy rate of  97.62\% in detecting DDoS attacks, with a further enhancement of approximately 5\% through transfer learning. Complementarily, DL combined with blockchain has been validated for malware detection in safety-critical autonomous vehicles, offering transferable design patterns for cyber-physical systems like HAPS \cite{patel2022_iwcmc_malware_av}.
\subsection{Access Controls}
Implementing robust access control mechanisms ensures that only authorized personnel can interact with sensitive systems and data, significantly reducing the risk of internal breaches. Strong identity authentication and fine-grained authorization policies are essential to protect HAPS ground segments, control links, and mission data from malicious attacks and unauthorized intrusions. In practice, attribute-based access control mechanisms can evaluate user identity, role, device attributes, and environmental context (e.g., time, location, network segment) to make dynamic access decisions across the communication infrastructure. Regular distribution of updated network state information and short-lived proxy credentials or session tokens enables efficient pre-authentication and limits the exposure window of compromised credentials, thereby strengthening the overall security posture of HAPS control and management planes \cite{nist_sp_800_162_abac}.

\subsection{Data Transmission Protection}
Recent work by various researchers has explored validated defenses against data poisoning during transmission attacks, with denoising techniques showing particular promise. However, finding a single defense strategy that is effective against all types of attacks remains a challenge. As wireless communication systems evolve, data transmission security becomes increasingly critical. The rapid emergence of novel adversarial attack techniques demands equally rapid defensive evolution. System breaches—whether in operational or simulated environments—can cause substantial damage, eroding user confidence and creating negative operational experiences~\cite{ilbert_breaking_2023}.

\subsection{Power Allocation Defense Mechanism}

Sahay et al. \cite{defending} demonstrated that adversarial denoising provides effective protection for HAPS power allocation systems against FGSM and PGD attacks \cite{goodfellow2014explaining, madry2017towards}, though this approach shows reduced effectiveness against more sophisticated Carlini \& Wagner (C\&W) attacks \cite{carlini2017towards}.

This performance difference arises from fundamental attack methodology differences. While FGSM and PGD generate detectable perturbations with large magnitudes and predictable statistical signatures, C\&W attacks employ sophisticated optimization that simultaneously minimizes adversarial loss and perturbation magnitude
. The resulting perturbations are nearly imperceptible and closely resemble natural signal variations, making them extremely difficult to distinguish from legitimate channel fading and interference. C\&W attacks can adaptively adjust to defense mechanisms, effectively bypassing static denoising thresholds and neutralizing statistical detection approaches. This limitation motivates the need for more advanced defense mechanisms, including certified defenses \cite{cohen2019certified}, adversarial training \cite{shafahi2019adversarial}, or ensemble methods that provide theoretical guarantees against adaptive attacks.

\subsection{Adversarial ML and Perception Defenses}
Learning-based perception and autonomy pipelines on HAPS and cooperating UAVs are susceptible to adversarial patch attacks. Outlier-Detection with Dimension Reduction (ODDR) couples robust dimensionality reduction with outlier scoring to reject patch-induced distribution shifts and has shown efficacy against patch-based perception attacks in autonomy stacks \cite{chattopadhyay2025_iccv_oddr}. Complementary techniques include adversarial training, input transformations, certified defenses, and model ensembling tailored for resource-constrained aerial platforms.

\subsection{ML-Based Attack Detection}
ML approaches for aerial network attack detection are attractive because they adapt to evolving threats. Recent deep-learning methods for network intrusion detection show promising results for UAVs and aerial communication systems \cite{shi2021generative}. In distributed settings, feature selection reduces communication and computation overhead for resource-constrained intrusion detection systems; Harris Hawks Optimization (HHO) has been demonstrated for secure Internet-of-Things (IoT) deployments \cite{hijazi2023_icc_hho}. However, HAPS deployments face unique constraints: limited on-board compute/energy and the absence of comprehensive, HAPS-specific attack datasets. Practical pipelines often rely on synthetic data or transfer learning from terrestrial networks, with potential performance degradation under stratospheric conditions due to different propagation characteristics and environmental factors. Temperature-induced hardware variation and increased cosmic radiation can further perturb ML inference. While ML-based detection is promising, most results target terrestrial or satellite settings; robust HAPS deployment will require HAPS-specific datasets, adaptation to stratospheric channels, and validation under representative environmental conditions.
\subsection{Software Security and Vulnerabilities Detection}
Software that integrates data and operational inputs from various hardware components is crucial for optimizing the performance and efficiency of HAPS. Such software is subject to vulnerabilities that could compromise critical operations and data integrity. Attackers might exploit these weaknesses to obtain unauthorized access, alter system operations, interfere with communication networks, or seize control of the HAPS platform. To reduce these risks, it is essential to implement strict software security practices: adopt secure coding guidelines, perform comprehensive vulnerability assessments and penetration testing, establish strong access controls, and promptly apply patches and updates to remediate known vulnerabilities.
 Wang et al. \cite{patchrnn} suggest a defense system named PatchRNN that uses deep learning to automatically identify confidential security patches in open-source software (OSS). The system utilizes descriptive keywords in the commit message and syntactic and semantic features at the source-code level. The system was assessed using a significant real-world patch dataset and a case study on NGINX. The findings show that the PatchRNN system can successfully identify concealed security patches with a minimal rate of false positives.
Thapa et al. \cite{thapa} investigate how large transformer-based language models can be utilized to identify software vulnerabilities in C/C++ source code by utilizing the transferability of knowledge acquired from natural language processing. The article introduces a structured approach for translating source code, preparing models, and making inferences, showcasing the effectiveness of transformer-based language models in detecting software vulnerabilities through analyzing empirical datasets. Fu et al. \cite{fu} suggest LineVul, a method that utilizes a Transformer-based model to forecast software vulnerabilities on a line-by-line basis. The method is tested on a real-world dataset containing over 188k C/C++ functions. It obtains a greater F1-measure for predictions at the function level and improved Top-10 accuracy for predictions at the line level in comparison with baseline methods. The examination also reveals that LineVul correctly forecasts vulnerable functions impacted by the top 25 most risky CWEs.

\subsection{Secure Boot}
Embedded devices within HAPS subsystems exchange data and coordinate operations, containing confidential information such as GPS location, telemetry data, and access keys, making them vulnerable to security risks.
\vspace{-.8cm}

\subsubsection{Stratospheric Environmental Challenges}
HAPS secure boot implementations must address extreme stratospheric conditions that challenge conventional cryptographic systems:

\textbf{Environmental Impact:} Stratospheric conditions impose extreme temperature ranges, low pressure, and significantly elevated cosmic radiation compared to terrestrial environments. Experimental radiation-beam studies of modern embedded processors (e.g., Xilinx UltraScale+ MPSoCs) show substantial susceptibility to single-event upsets (SEUs) in configuration memory, on-chip caches, and power-regulation subsystems \cite{anderson2018neutron}. These findings highlight the need for ECC-protected memory, fault-tolerant boot loaders, and radiation-hardened execution environments during the secure boot process.

\textbf{Performance Degradation:} RSA signature verification and AES decryption exhibit reduced throughput when subjected to temperature extremes and fluctuating supply conditions typical of high-altitude environments. As a result, secure boot sequences tend to be slower and more resource-intensive in stratospheric operations than at sea level, requiring optimizations such as hardware accelerators or pre-validated trust anchors.
\vspace{-.8cm}
\subsubsection{Multi-Stage Chain of Trust Architecture}

HAPS platforms implement a hierarchical chain of trust adapted for stratospheric deployment:

\textbf{Hardware Root of Trust:} TPM-based attestation architectures provide hardware-anchored key storage and secure measurement of early boot components. Recent TPM-enabled remote attestation frameworks such as EMBRAVE demonstrate scalable verification of embedded and IoT platforms under dynamic and distributed environments \cite{s25175514}. Dual-TPM configurations or redundant secure elements can further mitigate radiation-induced failures during high-altitude operation.

\textbf{Multi-Level Verification:} Enhanced verification chains include: (1) hardware authenticity checks using RSA-4096 or ECC P-384 keys, (2) immutable bootloader verification via SHA-3-256 digests, (3) kernel and hypervisor validation, (4) runtime attestation performed periodically during flight, and (5) continuous integrity monitoring to detect tampering or SEUs during long-duration missions.
\vspace{-.8cm}
\subsubsection{Implementation Challenges}

\textbf{Power and Performance Impact:} Secure boot verification increases power consumption during startup phases and adds significant time to startup sequences. Parallel verification pipelines can reduce overhead. Hardware Security Modules (HSMs) add latency per signature verification.

\textbf{Cost and Infrastructure:} Radiation-hardened cryptographic processors cost significantly more than commercial equivalents while providing reduced computational performance. Ground station costs increase substantially to support secure boot operations. Comprehensive secure boot may reduce mission duration for extended HAPS missions.

Zhang et al.\cite{zhang2022pa} present a processor authentication protocol called PA-Boot for verifying the legitimacy of application processors and securing interprocessor communication confidentiality in multiprocessor systems during the secure boot sequence.   
\subsection{Ground Station Defenses}
Ground stations provide centralized monitoring and real-time threat detection capabilities through cloud-based processing. They monitor onboard computer system (OBCS) activities where CPU cycle changes indicate potential malfunctions or attacks. For example, momentum wheel control failures cause decreased CPU activity as data collection stops, enabling ground operators to isolate malfunctions and maintain mission resilience through passive strategies and alternative actuator systems \cite{zhang2022pa}.

\subsection{Defense Mechanism Integration and Implementation Feasibility}

        \textbf{Cross-Subsystem Integration Requirements:} Defense mechanisms must operate cooperatively across communication, control, and power management subsystems. IDS monitoring in networked systems generates substantial daily log data requiring secure transmission, and DL-based IDS for IoT-scale traffic have demonstrated favorable detection/generalization trade-offs \cite{hamidouche2024_iot}. Power consumption increases during active defense operations, requiring energy-aware security algorithms \cite{shafique2021security}.

\textbf{Deployment Readiness Assessment:} Based on current technology maturity and ITU technical guidelines \cite{itu2019recommendation}, high readiness solutions include encryption, authentication, and basic IDS. Medium readiness solutions encompass frequency hopping and directional antennas. Lower readiness solutions involve advanced AI defenses and distributed consensus protocols requiring further research and development.

\textbf{Cost-Benefit Analysis:} Cybersecurity economics research indicates significant variation in implementation costs based on security level \cite{gordon2002economics}. Basic cybersecurity measures provide foundational protection with a moderate cost investment. Advanced mechanisms offer enhanced protection but require substantial financial commitment. Comprehensive implementation represents the highest protection level with corresponding cost implications. Recommended phased deployment prioritizes high-readiness mechanisms for immediate security improvements.

\section{Case Study: Experimental Validation through DDoS Attack Simulation on HAPS Platform}

This section provides experimental validation of our security analysis through a comprehensive DDoS attack simulation, demonstrating the practical impact of cyber threats on HAPS platforms and validating our proposed mitigation strategies.

We simulate a DDoS attack scenario on a HAPS system using OMNeT++ and the INET framework. Figure \ref{ddos_sim} illustrates the results where the x-axis represents time and the y-axis shows packet counts. The base station sends legitimate control commands to the HAPS (green line), while attackers inject malicious packets to overload the system (blue line). The total number of received packets (orange line) is tracked against the HAPS processing capacity (red dotted line). Once the total exceeds the allowed limit, the legitimate source can no longer deliver commands and the ground link is effectively severed. The simulation artifacts and code for this attack are available at \cite{simu}.

\subsection{Simulation Setup and Methodology}
We implemented a DDoS attack scenario using OMNeT++ simulation software with the INET framework to evaluate HAPS vulnerability under realistic attack conditions. The simulation models a stratospheric communication environment where:
\begin{itemize}
    \item Base stations send legitimate control commands to HAPS platforms (green line in Figure \ref{ddos_sim})
    \item Coordinated attackers inject malicious packets to overwhelm system resources (blue line)
    \item Total received packets are monitored against system capacity thresholds (orange line vs. red threshold)
\end{itemize}

\subsection{Experimental Results and Performance Analysis}
\begin{figure}[h]
    \centering
    \includegraphics[width=0.5\textwidth]{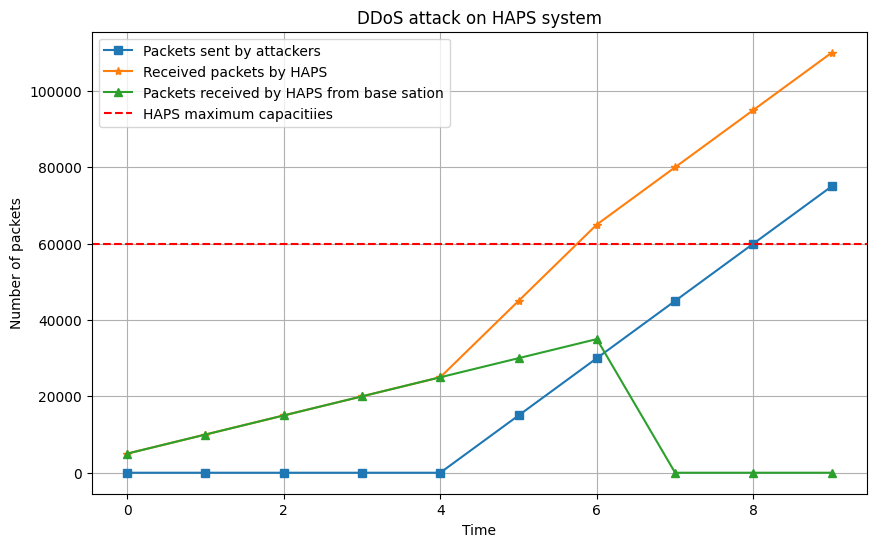}
    \caption{Simulation of DDoS Attack on HAPS Platform}
    \label{ddos_sim}
\end{figure}

Figure \ref{ddos_sim} presents our simulation results with time along the x-axis and packet volume on the y-axis. The experimental validation reveals several important findings:

\textbf{Attack Impact Assessment:} Our results show that when incoming packet volumes (orange line) exceed HAPS processing capacity (red dotted line), legitimate communications experience severe disruption, effectively severing ground control connections. This behavior validates our vulnerability analysis and underscores the necessity of the mitigation strategies we propose.

\textbf{Defense Mechanism Validation:} The simulation environment serves as a testbed for evaluating our proposed security measures:
\begin{itemize}
    \item Adaptive filtering performance in separating legitimate traffic from attack packets
    \item Intrusion detection system response times and accuracy under sustained attack
    \item Resource allocation strategy effectiveness during peak attack intensities
\end{itemize}

\textbf{Implementation Implications:} These experimental results offer concrete metrics for deployment planning, including packet processing thresholds, attack detection latency requirements, and system recovery timeframes essential for operational HAPS security implementations.

The case study demonstrates both HAPS vulnerability to DDoS attacks and the practical effectiveness of our proposed countermeasures. The simulation methodology and approach provide insights for future HAPS security research and implementation strategies.

\section{Conclusion}
This work examines the cybersecurity challenges facing HAPS, with particular attention to the privacy and security vulnerabilities inherent in these stratospheric platforms. As HAPS assume increasingly central roles in high-speed communication networks, their extensive coverage capabilities and stratospheric positioning create both opportunities and risks across sensing, machine-type communications, UAV coordination, and IoT applications.

Our analysis of HAPS subsystem architectures focused on identifying components critical for continuous operation, environmental adaptability, and reliable performance under extreme stratospheric conditions. The examination of HAPS network architectures and communication links reveals their pivotal role in connectivity enhancement, while simultaneously exposing emerging cybersecurity risks that threaten these systems and their dependent operations.

The stratospheric operating environment presents unique cybersecurity challenges not encountered in terrestrial or satellite systems. This research demonstrates the complexity of threats facing HAPS—from traditional jamming and spoofing to sophisticated intrusion attempts and power allocation attacks. Effective protection requires comprehensive approaches incorporating robust encryption, advanced intrusion detection, and intelligent response systems. Our integration of AI and machine learning capabilities enables real-time threat detection and response, significantly strengthening HAPS network security posture against both conventional and emerging attack vectors.

Future implementation of AI-based intrusion detection systems integrated directly into HAPS platforms will prove essential for refining defense mechanisms, ensuring these critical platforms can operate securely and efficiently as cyber threats continue evolving. The research presented here provides foundational security principles for the next generation of stratospheric communication infrastructure.

\vspace{0.2cm}
\noindent\textbf{Limitations:} The analysis and validation in this paper are simulation-based and rely on models and datasets representative of HAPS-like conditions rather than measurements from flight hardware. We do not report on-air experiments or hardware-in-the-loop evaluations, and we do not claim comprehensive coverage of all attack classes (e.g., sophisticated supply-chain compromises or coordinated multi-vector campaigns). Results should therefore be interpreted as indicative for design and planning. Future work will prioritize hardware-in-the-loop evaluation, controlled over-the-air tests, and broader cross-layer datasets collected under stratospheric conditions.

\vspace{0.2cm}
\noindent\textbf{Future Research Directions:} Several promising research avenues emerge from this work:
\begin{itemize}
    \item \textit{Stratospheric dataset development:} The community needs comprehensive, labeled datasets capturing RF/FSO signals, telemetry data, and network traces under realistic stratospheric conditions. Such datasets would enable proper training and evaluation of IDS and adversarial defense systems for actual deployment scenarios.
    \item \textit{Over-the-air adversarial robustness:} Current ML models struggle with real-world RF impairments. Developing over-the-air robust models using certified defenses or randomized smoothing that maintain accuracy despite Doppler effects, fading, and hardware nonlinearities at 18-50 km altitudes remains challenging.
    \item \textit{Scalable frequency agility:} Multi-HAPS constellations require coordinated FHSS with sophisticated synchronization and spectrum sharing. This includes adaptive hop patterns that provide provable jamming resistance while coexisting with satellite systems.
    \item \textit{Distributed security architectures:} HAPS fleets need bandwidth-efficient federated learning approaches for IDS deployment. Privacy-preserving aggregation methods resistant to model poisoning and Byzantine failures are particularly critical.
    \item \textit{Hardware supply chain verification:} Practical Trojan detection protocols for avionics hardware require standardized side-channel analysis across process, voltage, and temperature variations. Cryptographic provenance and secure boot mechanisms must accommodate HAPS maintenance cycles.
    \item \textit{Robust resource management:} Power and gateway scheduling algorithms need theoretical guarantees against optimization-based attacks like C\&W while operating under strict power constraints typical of stratospheric platforms.
    \item \textit{Security-aware cross-layer design:} Joint optimization across PHY/MAC/network layers with security constraints involves complex tradeoffs between detection latency, false positives, and energy consumption that require systematic investigation.
    \item \textit{Standards and formal verification:} The field needs advancement in formal methods for verifying safety and security properties of critical HAPS software and hardware, along with standardized incident reporting and inter-platform trust mechanisms.
\end{itemize}

\bibliographystyle{IEEEtran}  
\bibliography{ref}

\begin{IEEEbiography}[{\includegraphics[width=1in,height=1.25in,clip,keepaspectratio]{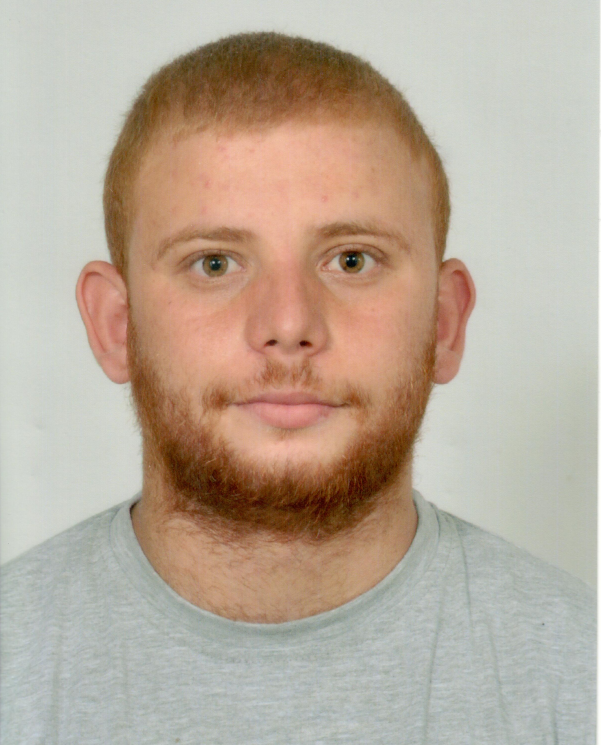}}]{Chaouki Hjaiji } (Student Member, IEEE) was born in Tunis, Tunisia. He has been pursuing an engineering degree at Tunisia Polytechnic School from 2022 to 2025. In 2025, he conducted a research internship at the Communication Systems Laboratory of EURECOM. He graduated with an engineering degree from Tunisia Polytechnic School in 2025.
\end{IEEEbiography}

\begin{IEEEbiography}[{\includegraphics[width=1in,height=1.25in,clip,keepaspectratio]{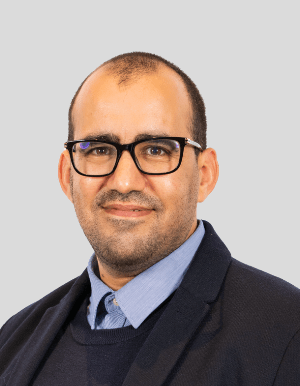}}]{Bassem Ouni } (Senior Member, IEEE) received his Ph.D. degree in computer science from the University of Nice Sophia Antipolis, Nice, France, in July 2013. From 2009 to 2013, he was a Lecturer with the University of Nice Sophia Antipolis (Polytech Nice Engineering School and Faculty of Sciences of Nice). From September 2013 to 2014, he was a Postdoctoral Fellow with Eurecom, Telecom ParisTech Institute, Sophia Antipolis, France. From 2015 to 2016, he was a Research Scientist with the Institute of Technology in Aeronautics, Space and Embedded Systems (IRT-AESE), Toulouse, France. From 2017 to 2018, he was a Lead Researcher with the Department of Electronics and Computer Science, University of Southampton, Southampton, U.K. From October 2018 to January 2022, he was a Lead Researcher with the French Atomic Energy Commission (CEA), LIST Institute, Paris, France, and an Associate Professor/Lecturer with the University of Paris Saclay and the ESME Sudria Engineering School, Paris. Also, he managed several industrial collaborations with ARM, Airbus Group Innovation, Rolls Royce, Thales Group, Continental, and Actia Automotive Group. He is currently a Lead Researcher with the Technology Innovation Institute, Abu Dhabi, United Arab Emirates. He co-authored many publications (book chapters, journals, and international conferences), and has organized and chaired several conferences and special sessions. He is an IEEE Senior member.
\end{IEEEbiography}

\begin{IEEEbiography}[{\includegraphics[width=1in,height=1.25in,clip,keepaspectratio]{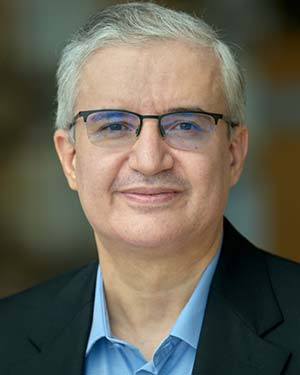}}]{Mohamed-Slim Alouini } (Fellow, IEEE) was born in Tunis, Tunisia. He earned his Ph.D. in Electrical Engineering from the California Institute of Technology (Caltech) in Pasadena, California, USA, in 1998. He has served as a faculty member at the University of Minnesota (UMN) in Minneapolis, Minnesota, USA, and Texas A\&M University at Qatar (TAMUQ) in Doha, Qatar. Since 2009, he has been a part of King Abdullah University of Science and Technology (KAUST) in Thuwal, Saudi Arabia, where he holds the position of Al-Khawarizmi Distinguished Professor in the Electrical and Computer Engineering Department. His research spans the design and performance analysis of diversity combining techniques, MIMO systems, multi-hop/cooperative communications, optical wireless communications, cognitive radio systems, green communication systems and networks, and integrated ground-airborne-space networks. He actively addresses global disparities in the distribution, access, and use of information and communication technologies, focusing on developing new generations of aerial and space networks to provide connectivity to remote, sparsely populated, and hard-to-reach areas. Dr. Alouini is a fellow of the World Wireless Research Forum (WWRF), the World Academy of Science (TWAS), OPTICA (formerly the Optical Society of America), and IEEE. He has co-received numerous awards across 26 journals, conferences, and challenges, with notable recognitions from IEEE and other international bodies. His editorial roles have included positions at IEEE Transactions on Communications, IEEE Transactions on Wireless Communications, IEEE Transactions on Mobile Computing, and the Wireless Communications and Mobile Computing Journal. Additionally, he has been the founding field chief editor of the Frontiers in Communications and Networks journal since 2020 and an editor for the IEEE Transactions on Aerospace and Electronics Systems since 2022.
\end{IEEEbiography}

\end{document}